\documentclass[12pt]{article}
\usepackage{amssymb,amscd,array}
\usepackage[active]{srcltx}
\catcode `\@=11
\@addtoreset{equation}{section}

\newtheorem{thm}{Theorem}[section]
\newtheorem{rem}[thm]{Remark}

\def\qed{\blacksquare}
\newcommand{\be}{\begin{equation}}
\newcommand{\ee}{\end{equation}}
\newcommand{\bea}{\begin{eqnarray}}
\newcommand{\eea}{\end{eqnarray}}
\newcommand{\R}{\mathbb{R}}
\newcommand{\N}{\mathbb{N}}
\newcommand{\C}{\mathbb{C}}

\begin{document}
\begin{titlepage}

\begin{center}
{\bf \Large{Third Order Anomalies in the Causal Approach  \\}}
\end{center}
\vskip 1.0truecm
\centerline{D. R. Grigore, 
\footnote{e-mail: grigore@theory.nipne.ro}}
\vskip5mm
\centerline{Department of Theoretical Physics,}
\centerline{Institute for Physics and Nuclear Engineering ``Horia Hulubei"}
\centerline{Bucharest-M\u agurele, P. O. Box MG 6, ROM\^ANIA}

\vskip 2cm
\bigskip \nopagebreak
\begin{abstract}
\noindent
We consider gauge models in the causal approach and study the third order of the perturbation theory. We are interested
in the computation of the anomalies in this order of the perturbation theory and for this purpose we analyse in detail
the causal splitting of the distributions with causal support relevant to tree and loop anomalies. 
\end{abstract}
%\newpage\setcounter{page}1
\end{titlepage}

\section{Introduction}

The most natural way to arrive at the Bogoliubov axioms of perturbative quantum field theory (pQFT) is by
analogy with non-relativistic quantum mechanics \cite{Gl}, \cite{H}. Suppose that we have a time-dependent interaction potential
$V$. Then one goes to the interaction picture and the time evolution is governed by the evolution equation:
\be
{d \over dt}U(t,s) = - i V_{\rm int}(t) U(t,s); \qquad U(s,s) = I.
\ee

This equation can be solved in some cases by a perturbative method, namely the series
\be
U(t,s) \equiv \sum {(-i)^{n}\over n!} \int_{\R^{n}} dt_{1} \cdots dt_{n} T(t_{1},\dots,t_{n})
\ee
makes sense. The operators 
$
T_{n}(t_{1},\dots,t_{n})
$
are called  {\it chronological products}; $n$ is called the 
{\it order} of the perturbation theory. They verify the following properties:

\begin{itemize}
\item 
Initial condition
\be
T_{1}(t) = V_{\rm int}(t)
\ee
\item 
Symmetry
\be
T_{n}(t_{\pi(1)},\dots,t_{\pi(n)}) = T_{n}(t_{1},\dots,t_{n})
\ee
for all permutations 
$
\pi
$
of
$
\{1,\dots,n\}.
$
\item 
Causality
\bea
T_{n}(t_{1},\dots,t_{n}) = T_{m}(t_{1},\dots,t_{m})~T_{n-m}(t_{m+1},\dots,t_{n}),
\nonumber\\
{\rm for} 
\quad t_{j} > t_{k}, \quad j = 1,\dots,m; k = m+1,\dots,n.
\eea
\item 
Unitary
\be
U(t,s)^{\dagger}~U(t,s) = I
\ee
 
In terms of the chronological products, define the {\it anti-chronological products} as follows: because of the 
symmetry property we can write
$
T(I) = T(i_{1},\dots,i_{k})
$
for
$
I = \{ i_{1},\dots,i_{k} \}.
$
Then the anti-chronological products are 
\be
(-1)^{n}~\bar{T}_{n}(t_{1},\dots,t_{n}) = \sum_{m=1}^{n} (-1)^{m} \sum_{I_{1},\dots,I_{m} \in part(I)}~ T_{I_{1}}\cdots T_{I_{m}}
\ee
where 
$
I_{1}, \dots, I_{m}
$
is a partition of $I$. The the unitarity axiom is equivalent to
\be
\bar{T}_{n}(t_{1},\dots,t_{n}) = T_{n}(t_{1},\dots,t_{n})^{\dagger}
\ee
\item 
Invariance properties

If the interaction potential is translation invariant then we have
\bea
T_{n}(t_{1} + \tau,\dots,t_{n} + \tau) = T_{n}(t_{1},\dots,t_{n}),\quad \forall \tau \in \R
\eea
\end{itemize}
We can write an explicit formula
\be
T_{n}(t_{1},\dots,t_{n}) = \sum \theta(t_{\pi(1)} - t_{\pi(2)})~\cdots~\theta(t_{\pi(n-1)} - t_{\pi(n)})~V_{int}(t_{1})~\cdots~V_{int}(t_{n}) 
\label{explicit}
\ee

The purpose is to generalize this idea in the relativistic context especially the causality property. Essentially
we try to substitute
$
t \in \R
$
by a Minkowski variable 
$
x \in \R^{4}.
$
The chronological operators will be some operators
$
T(x_{1},\dots,x_{n})
$
and all the preceding axioms can be easily generalized: the symmetry and the unitarity axioms remain unchanged and 
for the invariance axiom we have to substitute the translation group with Poincar\'e group. 
The causally axiom is more subtle. We have to replace
temporal succession
$
t_{1} > t_{2} 
$
by causal succession
$
x_{1} \succ x_{2}
$
which means that 
$
x_{1}
$
should not be in the past causal shadow of
$
x_{2}
$
i.e.
$
x_{2} \cap (x_{1} + \bar{V}^{+}) = \emptyset.
$
In formulas: if
$x_{i} \succ x_{j}, \quad \forall i \leq k, \quad j \geq k+1$
then we have:
\be
T(x_{1},\dots,x_{n}) =
T(x_{1},\dots,x_{k})~T(x_{k+1},\dots,x_{n}).
\label{causality1}
\ee
From here it follows that the ``initial condition"
$
T(x)
$
should satisfy
\be
[ T(x), T(y) ] = 0,\qquad (x - y)^{2} < 0
\ee
where for the Minkowski product we use the convention 
$
1,-1,-1,-1.
$
It is surprisingly difficult to obtain solutions of the preceding equation. The relevant solution for pQFT are in fact 
some distribution-valued operators, called Wick monomials. They can be associated to free fields (or generalized free
fields) and act in some Hilbert space of the Fock type. This is in accord with our intuition that in pQFT we
should be able to describe scattering processes with creation and annihilation of particles. However, in this case
the formula (\ref{explicit}) makes no sense. It involves an illegal operation: the multiplication of distributions.
It is better to try to solve directly the axioms of pQFT in an recursive way. 

So we start from Bogoliubov axioms \cite{BS}, \cite{EG} as presented in \cite{DF}; for every set of Wick polynomials 
$ 
A_{1}(x_{1}),\dots,A_{n}(x_{n}) 
$
acting in some Fock space
$
{\cal H}
$
one associates the operator-valued distributions
$ 
T^{A_{1},\dots,A_{n}}(x_{1},\dots,x_{n})
$  
called chronological products; it will be convenient to use another notation: 
$ 
T(A_{1}(x_{1}),\dots,A_{n}(x_{n})). 
$ 
The expression
$
T(x_{1},\dots,x_{n})
$
corresponds to the choice
$
A_{1} = \cdots A_{n} = T
$
and the generalization to the case of distinct
$
A_{1}, \cdots, A_{n}
$
is possible because the symmetry axioms suggests that a sort of polarization formula is possible. 

The axioms for the chronological products remain unchanged, only the symmetry axiom should be replaced by
skew-symmetry in all arguments: for arbitrary
$
A_{1}(x_{1}),\dots,A_{n}(x_{n})
$
we should have
\be
T(\dots,A_{i}(x_{i}),A_{i+1}(x_{i+1}),\dots,) =
(-1)^{f_{i} f_{i+1}} T(\dots,A_{i+1}(x_{i+1}),A_{i}(x_{i}),\dots)
\label{sqew}
\ee
where
$f_{i}$
is the number of Fermi fields appearing in the Wick monomial
$A_{i}$.

Even in the simplest case when the Fock space is generated by a real scalar field 
$
\phi(x)
$
and the interaction Lagrangian is a Wick monomial
$
T(x) = :\Phi^{4}(x):
$
the construction of the chronological products is a surprisingly difficult problem.

There are, at least to our knowledge, tree rigorous ways to do that; for completeness we remind them following \cite{ano-free}:

(a) {\it Hepp axioms} \cite{H}: one rewrites Bogoliubov axioms in terms of vacuum averages of chronological products
$
< \Omega, T^{A_{1},\dots,A_{n}}(x_{1},\dots,x_{n})\Omega>
$
(more precisely the contributions associated to various Feynman graph). One needs a regularization procedure for the 
Feynman amplitudes. Moreover, one proves that the renormalized Feynman amplitudes can be obtained from the formal
Feynman rules if one adds appropriate counterterms in the interaction Lagrangian.

(b) {\it Polchinski flow equations} \cite{P}, \cite{S}: one considers an ultra-violet cut-off 
$
\Lambda
$
for the Feynman amplitudes and establishes some differential equations (in this parameter) for these amplitudes.
The equations have such a structure that one can obtain the Feynman amplitudes by some recursive procedure and integration
of these differential equations. The computations are usually done in the Euclidean framework and is less obvious that the end 
result will verify Bogoliubov axioms.

(c) {\it The causal approach} due to Epstein and Glaser \cite{EG}, \cite{Gl}: is a recursive procedure for the basic objects
$ 
T(A_{1}(x_{1}),\dots,A_{n}(x_{n}))
$
and reduces the induction procedure to a distribution splitting of some distributions with causal support.  
In an equivalent way, one can reduce the induction procedure to the process of extension of distributions \cite{PS}. 
An equivalent point of view uses retarded products \cite{St1} instead of chronological products. The causal method is 
by far the most elementary, so we expect that this will stay true for more complicated models like gauge models.

In fact, a basic problem is the choice of the Fock space. Generally, we should consider some elementary particle described 
by some projective unitarity, irreducible representation of the Poincar\'e group, construct the associated Fock space
(taking into account the spin-statistics theorem) and build free fields as combinations of the creation and annihilation
operators. Because the irreducible representation of the Poincar\'e group are unique, up to an unitary transformation,
one would expect that the construction of the associated pQFT is also essentially unique. However, this is not so obvious.
For instance, the scalar particles are usually described by a scalar function
$
\Phi: \R^{4} \rightarrow \C
$
verifying Klein-Gordon equation. But they can also be described by a skew-symmetric tensor
$
t: \R^{4} \rightarrow \C^{4} \otimes \C^{4}
$
verifying Dirac equation in both entries \cite{V}, pg. 360. It is not obvious that if we work in this representation
for the scalar field we will obtain the same results as above. So it is a bit of art to choose a ``nice" concrete 
representation of the Hilbert space of an elementary particle. This task is more difficult for gauge theories 
which describe particles of higher spin. If we describe a particle of spin $1$ by a vector field and try to 
consider only the physical degrees of freedom (three for the massive case and only two for the massless case)
we end up with non-renormalizable theories. 

However, one can save renormalizablility using ghost fields. There are two ways to do that:

(A) In BRST approach one introduces even and odd Grassmann classical fields; the odd fields are the so-called {\it ghost} (or
Faddeev-Popov) fields. Then one can try to make sense of the formal path integral and ends up with some consistency relation - 
the master equation \cite{HT}. Presumably, if a solution of this equation can be found, one would be able to construct 
the chronological products with the desired properties, although a rigorous proof of this fact seems to be missing,
at least to our knowledge. A supplementary problem in the functional formalism is that the Green
functions are affected by infra-red divergences; an adiabatic limit must be
performed and as it can be seen from the paper of Epstein and Glaser, this
limit is not easy to perform. 

(A') A variant of the preceding idea is the use of the Zinn-Justin relation \cite{Z-J}.

(B) The {\it causal approach} of Scharf and collaborators \cite{Sc1}, \cite{Sc2}. In this approach one makes sense of the 
ghost fields as well defined fields in some mathematical Fock space with physical and non-physical states. One has to
select the physical states by a certain gauge condition and the chronological products should leave invariant the set
of physical states. 

We remind the details: the theories are defined in a Fock space
$
{\cal H}
$
with indefinite metric, and one selects the physical states assuming the existence of
an operator $Q$ called {\it gauge charge} which verifies
$
Q^{2} = 0
$
and such that the {\it physical Hilbert space} is by definition
$
{\cal H}_{\rm phys} \equiv Ker(Q)/Im(Q).
$
One assigns a {\it ghost number} to every field and this gives a grading in the Hilbert
space
$
{\cal H}
$
and in the space of Wick monomials in
$
{\cal H}.
$
If we consider that the gauge charge has ghost number $1$ then 
the graded commutator
$
d_{Q}
$
of the gauge charge with any operator $A$ of fixed grading number
\be
d_{Q}A = [Q,A]
\ee
makes sense and is raising the ghost number by a unit. It means that
$
d_{Q}
$
is a co-chain operator in the space of Wick polynomials. From now on
$
[\cdot,\cdot]
$
denotes the graded commutator.
 
A gauge theory assumes also that there exists a Wick polynomial of null ghost
number
$
T(x)
$
called {\it the interaction Lagrangian} such that
\be
~[Q, T] = i \partial_{\mu}T^{\mu}
\label{gau1}
\ee
for some other Wick polynomials
$
T^{\mu}.
$
This relation means that the expression $T$ leaves invariant the physical
states, at least in the adiabatic limit. Indeed, if this is true we have:
\be
T(f)~{\cal H}_{\rm phys}~\subset~~{\cal H}_{\rm phys}  
\label{gau2}
\ee
up to terms which can be made as small as desired (making the test function $f$
flatter and flatter). In all known models one finds out that there exists a
chain of Wick polynomials
$
T^{I}
$
(where $I$ is a collection of indexes
$
I = [\nu_{1},\dots,\nu_{p}],~p = 0,1,\dots
$
and the brackets emphasize the complete antisymmetry in these indexes) such that
\bea
T \equiv T^{\emptyset}
\nonumber\\
\omega(T^{I}) = \omega_{0},~\forall I
\nonumber\\
gh(T^{I}) = |I|
\eea
and we have
\be
d_{Q}T^{I} = i~\partial_{\mu}T^{I\mu}.
\label{descent1}
\ee

It is clear that we should have 
$
T^{I} = 0,~|I| > 4
$
but in the Yang-Mills case we have
$
T^{I} = 0,~|I| > 2.
$

Now we can construct the chronological products
\be
T^{I_{1},\dots,I_{n}}(x_{1},\dots,x_{n}) \equiv
T(T^{I_{1}}(x_{1}),\dots,T^{I_{n}}(x_{n}))
\label{chronos}
\ee
according to the recursive procedure. We say that the theory is gauge invariant
in all orders of the perturbation theory if the following set of identities
generalizing (\ref{descent1}):
\be
d_{Q}T^{I_{1},\dots,I_{n}} = 
i \sum_{l=1}^{n} (-1)^{s_{l}} {\partial\over \partial x^{\mu}_{l}}
T^{I_{1},\dots,I_{l}\mu,\dots,I_{n}}
\label{gauge}
\ee
are true for all 
$n \in \N$
and all
$
I_{1}, \dots, I_{n}.
$
Here we have defined
\be
s_{l} \equiv \sum_{j=1}^{l-1} |I|_{j}.
\ee
In particular, the case
$
I_{1} = \dots = I_{n} = \emptyset
$
it is sufficient for the gauge invariance of the scattering matrix, at least
in the adiabatic limit: we have the same argument as for relation (\ref{gau2}).

Such identities can be usually broken by {\it anomalies} i.e. expressions of the
type
$
A^{I_{1},\dots,I_{n}}
$
which are quasi-local and might appear in the right-hand side of the relation
(\ref{gauge}). In a previous paper we have emphasized the cohomological
structure of this problem \cite{sr3}. We consider a {\it cochain} to be
an ensemble of distribution-valued operators of the form
$
C^{I_{1},\dots,I_{n}}(x_{1},\dots,x_{n}),~n = 1,2,\cdots
$
(usually we impose some supplementary symmetry properties) and define the
derivative operator $\delta$ according to
\be
(\delta C)^{I_{1},\dots,I_{n}}
= \sum_{l=1}^{n} (-1)^{s_{l}} {\partial\over \partial x^{\mu}_{l}}
C^{I_{1},\dots,I_{l}\mu,\dots,I_{n}}.
\ee
We can prove that 
\be
\delta^{2} = 0.
\ee
Next we define
\be
s = d_{Q} - i \delta,\qquad \bar{s} = d_{Q} + i \delta
\ee
and note that
\be
s \bar{s} = \bar{s} s = 0.
\label{sbars}
\ee
We call {\it relative cocycles} the expressions $C$ verifying
\be
sC = 0
\ee
and a {\it relative coboundary} an expression $C$ of the form
\be
C = \bar{s}B.
\ee
The relation (\ref{gauge}) is simply the cocycle condition
\be
sT = 0.
\label{cocycle}
\ee

This cohomological structure is similar but different from the 
well-known cohomology of the BRS(T) operator \cite{BRS}. Our BRST 
operator $s$ is a linear operator so it make sense in a Hilbert space; 
the BRS(T) operator from \cite{BRS} 
is a non-linear operator acting on polynomials in the classical fields and 
their derivatives. In fact, formally, our BRST operator is the linear part of
the usual BRS expression.

If we can prove that this relation is valid up to the order 
$
n - 1
$
then in order $n$ this relation is valid up to {\it anomalies}:
\be
sT = {\cal A}
\ee
where the anomalies in the right hand side have the generic form
\be
{\cal A}(x_{1},\dots,x_{n}) = \sum p_{i}(\partial)\delta(x_{1},\dots,x_{n})~
W_{i}(x_{1},\dots,x_{n}).
\ee
Here 
\be
\delta(x_{1},\dots,x_{n}) 
= \delta(x_{1} - x_{n}) \cdots \delta(x_{n-1} - x_{n}),
\ee
$
p_{i}
$
are polynomials in the partial derivatives and 
$
W_{i}
$
are Wick polynomials. There is a bound on the number
\be
deg({\cal A}) \equiv supp_{i}~\{deg(p_{i}) + \omega(W_{i})\}
\ee
coming from the power counting theorem; here 
$
deg(p)
$
is the degree of the polynomial $p$ and
$
\omega(W)
$
is the canonical dimension of the Wick polynomial $W$. We call this number
the {\it canonical dimension of the anomaly}. For instance if the interaction
Lagrangian and the associated expressions
$
T^{I}
$
verify
$
\omega(T^{I}) = 4
$
(as is the case of Yang-Mills models) then the canonical dimension of the
anomaly is 
$
\leq 5
$. 
The contributions corresponding to maximal degree will be called {\it dominant}. 

We note that from (\ref{cocycle}) it follows that the anomaly must verify a
consistency relation of the Wess-Zumino type
\be
\bar{s}{\cal A} = 0.
\label{WZ}
\ee

Such type of relations have intensively used to obtain the generic form of the 
anomalies in the causal approach in \cite{cohomology}. 

According to our knowledge, there is no rigorous proof of the equivalence between the functional formalism and 
the causal formalism which we use here. 

A systematic study for the loop contributions in the third order of the perturbation theory in the causal
approach appears in \cite{loop}. In this paper we consider the Yang-Mills models up to the third order studying all 
contributions: tree and loop; for the loop anomalies we present a simplified version. The basic idea is
to isolate some typical numerical distributions with causal support appearing
in the loop contributions in the second and the third order of the
perturbation theory; then we prove that some identities verified by these
distributions can be causally split without anomalies. This idea is in the
spirit of the master Ward identity considered in the literature \cite{DB},
\cite{DF1}, but the actual proof of our identities seems to be considerably
different. 

In the next Section we will give a minimal account of the gauge
theories in the causal approach. Next we turn to the one-loop anomalies in the second 
and third order of perturbation theory in Sections \ref{second} and \ref{third}.

\newpage

\section{General Gauge Theories\label{ggt}}
\subsection{Perturbation Theory}
 
The axioms of perturbation theory of pQFT in the Bogoliubov framework have been described in the introduction; for more
details see \cite{algebra}. We only remind two supplementary axioms.

(a) {\it Wick expansion property} 

It can be proved \cite{EG} that this system of axioms can be supplemented with
\bea
T(A_{1}(x_{1}),\dots,A_{n}(x_{n}))
\nonumber \\
= \sum \quad
<\Omega, T(A^{\prime}_{1}(x_{1}),\dots,A^{\prime}_{n}(x_{n}))\Omega>~~
:A^{\prime\prime}_{1}(x_{1}),\dots,A^{\prime\prime}_{n}(x_{n}):
\label{wick-chrono2}
\eea
where
$A^{\prime}_{i}$
and
$A^{\prime\prime}_{i}$
are Wick submonomials of
$A_{i}$
such that
$A_{i} = :A^{\prime}_{i} A^{\prime\prime}_{i}:$
and appropriate signs should be included if Fermi fields are present; here
$\Omega$
is the vacuum state. 

(b) {\it Power counting bound}

The order of singularity 
$
\omega(d)
$
of a distribution 
$d(x) \in {\cal S}^{\prime}(\R^{n})$ 
is defined in \cite{EG} (and slightly differently in \cite{St1}); 
essentially the Fourier transform 
$
\tilde{d}(p)
$
behaves for large momenta as
$
|p|^{\omega(d)}.
$

We can also include in the induction hypothesis a limitation on the order of
singularity of the vacuum averages of the chronological products associated to
arbitrary Wick monomials
$A_{1},\dots,A_{n}$;
explicitly:
\be
\omega(<\Omega, T^{A_{1},\dots,A_{n}}(x_{1},\dots,x_{n})\Omega>) \leq
\sum_{l=1}^{n} \omega(A_{l}) - 4(n-1)
\label{power}
\ee
where by
$\omega(d)$
we mean the order of singularity of the (numerical) distribution $d$ and by
$\omega(A)$
we mean the canonical dimension of the Wick monomial $A$.  The contributions
saturating the inequality (i.e. corresponding to the equal sign) will
be called {\it dominant}; they will produce dominant anomalies.

Up to now, we have defined the chronological products only for self-adjoint Wick
monomials 
$
A_{1},\dots,A_{n}
$
but we can extend the definition for arbitrary  Wick polynomials by linearity.

One can modify the chronological products without destroying the basic property
of causality {\it iff} one can make
\bea
T(A_{1}(x_{1}),\dots,A_{n}(x_{n})) \rightarrow  
\nonumber\\
T(A_{1}(x_{1}),\dots,A_{n}(x_{n})) +
\sum P_{j}(\partial)\delta(x_{1} - x_{n})\cdots\delta(x_{n-1} - x_{n})~W_{j}(x_{1},\dots,x_{n})
\label{renorm}
\eea
with 
$P_{j}$ 
monomials in the partial derivatives and
$
W_{j}
$
are Wick monomials. Some restrictions are following from power counting, Lorentz covariance and unitarity.

From now on we consider that we work in the four-dimensional Minkowski space and
we have the Wick polynomials
$
A, B,
$
etc. such that we have
\be
A(x)~B(y) = (-1)^{|A||B|}~B(y)~A(x),~\forall~x \sim y
\label{graded-comm}
\ee
i.e. for 
$
x - y
$
space-like these expressions causally commute in the graded sense. The
chronological products 
$
T(A_{1}(x_{1}),\dots,A_{n}(x_{n}))
$
are constructed according recursively using the causal commutators.

The basic recursive idea of Epstein and Glaser starts from the chronological products
$$ 
T(A_{1}(x_{1}),\dots,A_{m}(x_{m})) \quad m = 1,2,\dots
$$
up to order 
$
n -1
$
and constructs a causal commutator in order $n$. 
For instance for
$
n = 2
$
the {\it causal commutator} according to:
\be
D(A(x),B(y)) = A(x)~B(y) - (-1)^{|A||B|}~B(y)~A(x)
\ee
and after the operation of causal splitting one can obtain the second order 
chronological products. Generalizations of this formula are available for higher 
orders of the perturbation theory. In particular we have in the third order
\bea
D(A(x), B(y);C(z)) \equiv - [ \bar{T}(A(x), B(y)), C(z)]
\nonumber\\
+ (-1)^{|B||C|} [ T(A(x), C(z)), B(y)] + (-1)^{|A|(|B|+|C|)}  [ T(B(y), C(z)), A(x)]
\label{Dcausal}
\eea
where all commutators are understood to be graded.

\newpage
\subsection{Gauge Theories\label{ym}}
We will be interested in the following by Yang-Mills models; by this we mean the 
most general interaction between particles of spin 
$
0, 1/2
$
and $1$. The fields of spin $1$ are described using ghost fields and a suitable gauge operator. 
The Hilbert space of the model is generated by quantum free fields associated to the following types of particles:

1. Particles of null mass and helicity $1$ (photons and gluons). They are described by the vector fields 
$
v^{\mu}_{a}
$ 
(with Bose statistics) and the scalar fields 
$
u_{a}, \tilde{u}_{a}
$
(with Fermi statistics):
$
a \in I_{1}.
$

2. Particles of positive mass and spin $1$ (heavy Bosons). They are described by the vector fields 
$
v^{\mu}_{a}
$ 
(with Bose statistics) and the scalar fields 
$
u_{a}, \tilde{u}_{a}
$
(with Fermi statistics) and scalar fields
$
\Phi_{a}
$ 
with Bose statistics:
$
a \in I_{2}.
$

3. Scalar particles (the Higgs particle) 
$
\Phi_{a}
$ 
with Bose statistics:
$
a \in I_{3}.
$

4. Dirac fields 
$
\psi_{A}
$
with Fermi statistics:
$
A \in I_{4}.
$

To describe completely the model we need to give the following elements:

- The $2$-point functions; then we can generate the $n$-point functions using as a guide Wick theorem.

- A Hermiticity structure. 

All these elements can be found in preceding publications for instance \cite{cohomology}. One can use
the formalism described there to obtain in an unique way the expression of the interaction Lagrangian
$T$: it is (relatively) cohomologous to a non-trivial co-cycle of the form:
\bea
T = f_{abc} \left( {1\over 2}~v_{a\mu}~v_{b\nu}~F_{c}^{\nu\mu}
+ u_{a}~v_{b}^{\mu}~\partial_{\mu}\tilde{u}_{c}\right)
\nonumber \\
+ f^{\prime}_{abc} [\Phi_{a}~(\partial^{\mu}\Phi_{b} - m_{b} v_{b}^{\mu})~v_{c\mu} +
m_{b}~\Phi_{a}~\tilde{u}_{b}~u_{c}]
\nonumber \\
+ {1\over 3!}~f^{\prime\prime}_{abc}~\Phi_{a}~\Phi_{b}~\Phi_{c}
+ j^{\mu}_{a}~v_{a\mu} + j_{a}~\Phi_{a}.
\label{Tint}
\eea

The first line give the pure Yang-Mills 
interaction, the second line is the vector-scalar interaction, then comes the pure scalar interaction and the last two
terms give the interaction of the Dirac fields with the vector and resp. scalar particles mediated by the vector and scalar currents
\bea
j_{a}^{\mu} = \sum_{\epsilon = \pm}~
\overline{\psi} t^{\epsilon}_{a} \otimes \gamma^{\mu}\gamma_{\epsilon} \psi
\qquad
%\nonumber \\
j_{a} = \sum_{\epsilon = \pm}~
\overline{\psi} s^{\epsilon}_{a} \otimes \gamma_{\epsilon} \psi
\qquad
\gamma_{\epsilon} = {1\over 2}(I + \epsilon \gamma_{5}).
\label{current}
\eea
Here
$
t_{a} = (t_{a})_{AB}, \quad 
s_{a} = (s_{a})_{AB}
$
are matrices of dimension
$
I_{4}, (A, B \in I_{4})
$
and we group the Dirac fields in a vector column
$
\psi = (\psi)_{A}, (A \in I_{4}).
$
The expression $T$ above is constrained by Lorentz invariance and the bound $\leq 4$ on the canonical dimension. 
One can also give explicit formulas for the associated expressions
$
T^{\mu}, T^{\mu\nu}
$
(see \cite{cohomology}).

There are some linear relations fulfilled by the coefficients from (\ref{Tint}). We mention only the fact that
$
f_{abc}
$
is completely antisymmetric and that
$
f^{\prime}_{abc}
$
is antisymmetric in 
$
a, b.
$
\newpage
\subsection{Distributions with Causal Support and Causal Splitting}

We will use many times the so-called {\it central splitting of
causal distributions} \cite{Sc2}. We remind the reader the basic formula. Let 
$
d \in ({\cal S}^{4n})^{\prime}
$
be a distribution in the variables
$
x_{1},\dots,x_{n}
$
from the Minkowski space. Suppose that $d$ has causal support i.e.
\be
supp(d) \in \{(x_{1},\dots,x_{n}) | x_{j} - x_{n} \in V^{+} \cup V^{-}, j =
1,\dots,n - 1\}
\ee
and has the order of causality 
$
\omega = \omega(d) \in \N;
$
essentially this means that the Fourier transform 
$
\tilde{d}
$
of $d$ behaves for large momenta as
$
p^{\omega}.
$
It is a standard theorem in distribution theory that we can split
\be
d = a - r
\ee
where
\bea
supp(a) \in \{(x_{1},\dots,x_{n}) | x_{j} - x_{n} \in V^{+}, j = 1,\dots,n - 1\}
\nonumber \\
supp(r) \in \{(x_{1},\dots,x_{n}) | x_{j} - x_{n} \in V^{-}, j = 1,\dots,n - 1\}
\eea
are called the {\it advanced} and resp. {\it retarded} components of $d$; moreover, the splitting does not increases the order
of singularity. If
$
\omega(d) < 0
$
then $a$ and $r$ are uniquely determined; formally we have 
\bea
a(x) = \theta^{+}(x)~d(x)
\nonumber \\
r(x) = \theta^{-}(x)~d(x)
\eea
where
$
\theta^{\pm}
$
are some Heaviside functions separating the two pieces of the light cone. Let
us suppose that 
$
0 \not\in supp(\tilde{d});
$
then taking the Fourier transform we get for:
\be
\tilde{a}(p) = {i \over 2\pi} \int_{-\infty}^{\infty} dt 
{\tilde{d}(t p) \over 1 - t + i0},\qquad p \in V^{+} \cup V^{-}
\label{central1}
\ee
and the integral is convergent. If 
$
\omega(d) \geq 0
$
then the integral is not convergent any more and (as for the subtracted Cauchy
formula) we have:
\be
\tilde{a}(p) = {i \over 2\pi} \int_{-\infty}^{\infty} dt 
{\tilde{d}(t p) \over (t - i0)^{\omega}~(1 - t + i0)}
\label{central2}
\ee
and the integral is again convergent. This is the so-called 
{\it central} solution of the splitting problem. The general solution
is given by adding a polynomial in $p$ of maximal degree
$
\omega(d).
$
\newpage

\section{Second Order Causal Distributions\label{second}}

In second order we have some typical distributions. 
We remind the fact that the Pauli-Villars distribution is defined by
\be
D_{m}(x) = D_{m}^{(+)}(x) + D_{m}^{(-)}(x)
\ee
where 
\be
D_{m}^{(\pm)}(x) = \pm {i \over (2\pi)^{3}}~
\int dp e^{i p\cdot x} \theta(\pm p_{0}) \delta(p^{2} - m^{2})
\ee
such that
\be
D^{(-)}(x) = - D^{(+)}(- x).
\ee

This distribution has causal support. In fact, it can be causally split
(uniquely) into an
advanced and a retarded part:
\be
D = D^{\rm adv} - D^{\rm ret}
\ee
and then we can define the Feynman propagator and anti-propagator
\be
D^{F} = D^{\rm ret} + D^{(+)}, \qquad \bar{D}^{F} = D^{(+)} - D^{\rm adv}.
\ee
All these distributions have singularity order
$
\omega(D) = -2
$.

These distributions do appear in the tree contributions to the chronological products. One can have anomalies due to the 
following fact. From the gauge invariance (\ref{descent1}) we can prove that
\bea
sD(T^{I}(x), T^{J}(y)) \equiv 
\nonumber\\
d_{Q}D(T^{I}(x), T^{J}(y)) 
- i~\partial_{\mu}^{1}D(T^{I\mu}(x), T^{J}(y)) - (-1)^{|I|}~i~\partial_{\mu}^{2}D(T^{I}(x), T^{J\mu}(y)) = 0.
\eea
Use must be made of the Klein-Gordon equation
\be
(\square + m^{2})~D_{m}(x) = 0. 
\label{KG}
\ee
Indeed, we have to find the terms from
$
D(T^{I\mu}(x), T^{J}(y))
$
having a factor 
$
\partial^{\mu}_{1}D(x - y)
$
and the terms from
$
D(T^{I}(x), T^{J\mu}(y))
$
having a factor 
$
\partial^{\mu}_{2}D(x - y)
$
and we must use the Klein-Gordon equation from above to eliminate some terms. However, if we use the causal splitting and replace
$
D_{m}(x) \rightarrow D^{adv, ret, F}_{m}(x)
$
in the causal commutator, we are faced with the fact that the Klein-Gordon equation cannot be causally split: we have
\be
(\square + m^{2})~D^{F}_{m}(x) = \delta(x).
\label{split-KG}
\ee
These anomalies have been investigated in detail: see \cite{caciulata4} and references quoted there; in this reference we have
used an alternative method, namely the {\it off-shell} analysis. The main result is that 
the gauge invariance at the second-order tree level can be restored if one redefines the chronological products
$
T(A(x), B(y)) \rightarrow T(A(x), B(y)) + \delta(x - y)~N^{A,B}(x)
$
where the Wick polynomials 
$
N^{A,B}(x)
$
can be obtained from the quadri-linear terms of the classical Yang-Mills Lagrangian with the classical fields replaced by quantum fields 
and afterwards Wick ordering is applied. To be able to perform such a redefinition of the chronological products some bilinear relations 
must be obeyed by the coefficients from (\ref{Tint}). We mention only the fact that: (a)
$
f_{abc}
$
should fulfill the Jacobi identity, so they are the structure constants of some Lie algebra; (b) 
$
(f^{\prime}_{a})_{bc} \equiv - f^{\prime}_{bca}
$
and
$
t^{\epsilon}_{a}
$
should be representations of the Lie algebra emerging above; (c)
$
s^{\epsilon}_{a}
$
are tensor operators. We give below the expressions for the finite renormalizations of the chronological products:
\bea
N^{\emptyset\emptyset} = {i \over 2}~f_{abe}~f_{cde}~v_{a}^{\mu}v_{b}^{\nu}v_{c\mu}v_{d\nu}
- i~f^{\prime}_{eab}~f^{\prime}_{ecd}~\Phi_{a}v_{b\mu}\Phi_{c}v_{d}^{\mu}
+ {i \over 24}~\sum_{a,b,c,d}~g_{abcd}~\Phi_{a}\Phi_{b}\Phi_{c}\Phi_{d}
\nonumber\\
N^{[\mu]\emptyset} = - i~f_{abe}~f_{cde}~u_{a}v_{b}^{\nu}v_{c\nu}v_{d}^{\mu}
- i~f^{\prime}_{eab}~f^{\prime}_{ecd}~\Phi_{a}u_{b}\Phi_{c}v_{d}^{\mu}
\nonumber\\
N^{[\mu][\nu]} = - i~f_{abe}~f_{cde}~u_{a}v_{b}^{\nu}u_{c}v_{d}^{\mu}
\nonumber\\
N^{[\mu\nu]\emptyset} = - {i \over 2}~f_{abe}~f_{cde}~u_{a}u_{b}v_{c}^{\mu}v_{d}^{\nu}
\nonumber\\
N^{[\mu\nu][\rho]} = 0
\label{N}
\eea
where
\bea
g_{abcd} = F_{\{abcd\}}
\nonumber\\
F_{abcd} \equiv  \cases{ {1\over m_{a}}~{\cal S}_{bcd} \Bigl(f^{\prime}_{eba}~f^{\prime\prime}_{ecd} \Bigl) & if $ a \in I_{2}$ \cr
0 & if $a \in I_{1} \cup I_{3}$ \cr}.
\eea

For  one-loop contributions in the second order we need the basic distributions
\be
d_{D_{1},D_{2}}(x) \equiv
{1 \over 2}~[ D_{1}^{(+)}(x)~D_{2}^{(+)}(x) - D_{1}^{(-)}(x)~D_{2}^{(-)}(x) ]
\label{d2}
\ee
where
$
D_{j} = D_{m_{j}}
$
which also with causal support. This expression is linear in
$
D_{1}
$
and
$
D_{2}
$.
We will also use the notation
\be
d_{12} \equiv d(D_{1},D_{2}) \equiv d_{D_{1},D_{2}}
\ee
and when no confusion about the distributions
$
D_{j} = D_{m_{j}}
$
can appear, we skip all indexes altogether. The causal split
\be
d_{12} = d_{12}^{adv} - d_{12}^{ret}
\ee
is not unique because
$
\omega(d_{12}) = 0
$
so we make the redefinitions
\be
d_{12}^{adv(ret)}(x) \rightarrow d_{12}^{adv(ret)}(x) + c~\delta(x)
\ee
without affecting the support properties and the order of singularity.
The corresponding Feynman propagators can be defined as above and will be
denoted as
$
d_{12}^{F}
$.
Another way to construct them is to define for
$
x \not= 0
$
the distribution
\be
d^{(0)}_{12}(x) \equiv {1\over 2}~D_{1}^{F}(x)~D_{2}^{F}(x)
\ee
and to extend it to the whole domain using a standard result in distribution
theory (see the preceding Section). 

We will consider the case
$
D_{1} = D_{2} = D_{m}
$
and determine its Fourier transform; by direct computations it can be obtained
that
\be
\tilde{d}_{m,m}(k)
\equiv {1 \over (2\pi)^{2}} \int dx~ e^{i k\cdot x} d_{m,m}(x)
= - {1 \over 8 (2\pi)^{3}}~\varepsilon(k_{0})~\theta(k^{2} - m^{2}) 
\sqrt{1 - {4 m^{2} \over k^{2}}}.
\label{d-mm}
\ee

We also define the distributions
\bea
d^{\mu\nu}(x) = 
D^{(+)}_{m}(x) \partial^{\mu}\partial^{\nu}D^{(+)}_{m}(x) 
- D^{(-)}_{m}(x) \partial^{\mu}\partial^{\nu}D^{(-)}_{m}(x) 
\nonumber \\
f^{\mu\nu}(x) = 
\partial^{\mu}D^{(+)}_{m}(x) \partial^{\nu}D^{(+)}_{m}(x) 
- \partial^{\mu}D^{(-)}_{m}(x) \partial^{\nu}D^{(-)}_{m}(x) 
\eea

Performing a Fourier transform we can obtain the formula
\be
d^{\mu\nu}(x) = {2\over 3} 
\left(\partial^{\mu}\partial^{\nu} - 
{1\over 4}\eta^{\mu\nu}\square\right)d_{m,m}(x)
- {2 m^{2}\over 3}
(\partial^{\mu}\partial^{\nu} - \eta^{\mu\nu}\square)
d^{\prime}_{m,m}(x)
\ee
where we define the distribution
$
d^{\prime}_{m,m}(x)
$
through its Fourier transform:
\be
\tilde{d^{\prime}}_{m,m}(k) = {1\over k^{2}}~\tilde{d}_{m,m}(k).
\ee
This distribution also has causal support and it verifies
\be
\square d^{\prime}_{m,m} = - d_{m,m}.
\label{d-prime}
\ee
It can be proved that the central causal splitting preserves this relation. The
distribution
\be
f^{\mu\nu} = 2 {\cal D}_{1}^{\mu}{\cal D}_{2}^{\nu}d
\ee
is simply obtained as
\be
f^{\mu\nu} = \partial^{\mu}\partial^{\nu}d_{m,m} - d^{\mu\nu}.
\ee

The dominant contribution can produce anomalies of canonical dimension $5$ and
the super-renormalizable contributions can produce anomalies of canonical
dimension at most $3$. We investigate the dominant anomaly.

We now consider the one-loop contributions 
$
D_{(1)}^{IJ}(x,y)
$
from
$
D^{IJ}(x,y)
$
and we write for every mass $m$ in the game
\be
D_{m} = D_{M} + ( D_{M} - D_{m})
\label{split}
\ee
In this way we split 
$
D_{(1)}^{IJ}(x,y)
$
into a dominant contribution 
$
D_{\rm dominant}^{IJ}(x,y)
$
where everywhere
$
D_{m} \mapsto D_{M}
$
and a contribution where at least one factor
$
D_{m}
$
is replaced by the difference
$
D_{m} - D_{M}
$.
Because we have
\be
\omega(D_{m} - D_{M}) = -4
\ee
the second contribution will be super-renormalizable. The dominant contribution can produce anomalies of
maximal dimension 
$
\omega({\cal A}) = 5
$
and rest will produce anomalies with canonical dimension 
$
\omega({\cal A}) = 3.
$

We now consider the dominant contribution. By direct computations we obtain
\be
D_{\rm dominant}^{[\mu\nu]\emptyset}(x,y) = 0
\ee
\be
D_{\rm dominant}^{[\mu][\nu]}(x,y) = (\partial^{\mu}\partial^{\nu} - \eta^{\mu\nu}
\square)d_{M,M}(x - y) 
\tilde{g}_{ab} u_{a}(x) u_{b}(y)
\ee
\bea
D_{\rm dominant}^{[\mu] \emptyset}(x,y) = (\partial^{\mu}\partial^{\nu} -
\eta^{\mu\nu}
\square) d_{M,M}(x - y) 
\tilde{g}_{ab} u_{a}(x) v_{b\nu}(y)
\nonumber \\
+ \partial_{\nu} d_{M,M}(x - y) g_{ab} [ F^{\mu\nu}_{a}(x) u_{b}(y) - u_{a}(x)
F^{\mu\nu}_{b}(y) ]
\eea
\be
D_{\rm dominant}^{\emptyset [\mu]}(x,y) = - D_{(1)0}^{[\mu]\emptyset}(y,x)
\ee
\bea
D_{\rm dominant}^{\emptyset \emptyset}(x,y) = (\partial^{\mu}\partial^{\nu} -
\eta^{\mu\nu}
\square)d_{M,M}(x - y) 
\tilde{g}_{ab} v_{a\mu}(x) v_{b\nu}(y)
\nonumber \\
+ \partial_{\mu} d_{M,M}(x - y)
g_{ab} [ - F^{\mu\nu}_{a}(x) v_{b\nu}(y) +  \partial^{\mu}\tilde{u}_{a}(x)
u_{b}(y)
 + v_{a\nu}(x) F^{\mu\nu}_{b}(y) - u_{a}(x)  \partial^{\mu}\tilde{u}_{b}(y) ]
\nonumber \\
- d_{M,M}(x - y)g_{ab} F^{\mu\nu}_{a}(x) F_{b\mu\nu}(y) 
\nonumber \\
+ \partial_{\mu} d_{M,M}(x - y)
g^{(3)}_{ab} [ \Phi_{a}(x) \partial^{\mu}\Phi_{b}(y) -
\partial^{\mu}\Phi_{a}(x) \Phi_{b}(y) ]
- 2 d_{M,M}(x - y) g^{(3)}_{ab} \partial^{\mu}\Phi_{a}(x)
\partial_{\mu}\Phi_{b}(y)
\nonumber \\
- i \partial_{\mu} d_{M,M}(x - y)
[ \bar{\Psi}(x) A_{\epsilon} \otimes \gamma^{\mu}\gamma_{\epsilon}\Psi(y) 
- \bar{\Psi}(y) A_{\epsilon} \otimes \gamma^{\mu}\gamma_{\epsilon}\Psi(x)]
\nonumber \\
 + \square  d_{M,M}(x - y) g^{(4)}_{ab} \Phi_{a}(x) \Phi_{b}(y)
\eea
where we have defined some bilinear combinations in the constants appearing in
the 
interaction Lagrangian:
\bea
g_{ab} = f_{pqa} f_{pqb}
\qquad
g^{(1)}_{ab} = f^{\prime}_{pqa} f^{\prime}_{pqb} 
\qquad
g^{(2)}_{ab} = \sum_{\epsilon} Tr(t^{\epsilon}_{a} t^{\epsilon}_{b})
\qquad
g^{(3)}_{ab} = f^{\prime}_{apq} f^{\prime}_{bpq} 
\nonumber \\
g^{(4)}_{ab} = 2 \sum_{\epsilon} Tr(s^{\epsilon}_{a} s^{- \epsilon}_{b})
\qquad
\tilde{g}_{ab} \equiv {1\over 3}~(2 g_{ab} + g^{(1)}_{ab} + 4 g^{(2)}_{ab})
\qquad
A_{\epsilon} = \sum_{a} ( 2 t^{\epsilon}_{a} t^{\epsilon}_{a} +
s^{-\epsilon}_{a} s^{\epsilon}_{a}).
\eea

It is easy to see that the substitution
\be
d_{M,M}(x - y) \rightarrow d_{M,M}^{F}(x - y) 
\ee
gives the dominant contribution to the chronological product and does not produce anomalies.
So only anomalies of lower dimension can appear.

\newpage
\section{Third Order Causal Distributions\label{third}}

We must start from (\ref{Dcausal}) and use the complete formula for the second order causal products. 
Generically we have
\bea
T(A(x),B(y)) = :A(x)B(y): + T_{(0)}(A(x),B(y)) + T_{(1)}(A(x),B(y)) + T_{(2)}(A(x),B(y))
\nonumber\\
+ \delta(x - y) N^{A,B}(x)
\label{T2ren}
\eea
where the contributions 
$
T_{(l)},~l = 0,1,2
$
are the tree, one-loop and two-loops contributions and the last term is the finite
renormalization which must be done to eliminate the anomalies coming from the tree contributions - see (\ref{N}).
The two-loop term 
$
T_{(2)}
$
from (\ref{T2ren}) does not contribute to the causal commutator (\ref{Dcausal}) because it is a $c$-number.

We remain with two distinct types of terms in (\ref{Dcausal}): tree and loop graphs.

\subsection{Tree Graphs}

The first possibility is to consider the first term from the preceding formula of the generic form
\be
:A(x)B(y): = \sum :a_{1}(x)a_{2}(x)a_{3}(x)b_{1}(y)b_{2}(y)b_{3}(y):
\label{uncontracted}
\ee
When we commute this operator with
$
C(z) = :c_{1}(z)c_{2}(z)c_{3}(z):
$
we can take a contraction of a factor $a$ with a factor $c$ and a contraction of a factor $b$ with another factor $c$.

Another possibility comes from the second term of (\ref{T2ren}) with the generic form
\be
T_{(0)} (A(x),B(y)) = \sum p_{j}(\partial)D^{F}_{m}(x - y)~:a_{1}(x)a_{2}(x)b_{1}(y)b_{2}(y):
\label{T0}
\ee
When we commute this operator with
$
C(z) = :c_{1}(z)c_{2}(z)c_{3}(z):
$
one possibility is to contract one of the factors $a$ (or one of the factors $b$) with a factor $c$. 

These relevant causal distributions are: 
\bea
d^{(1)}_{D_{1},D_{2}}(x,y,z) \equiv 
\bar{D}^{F}_{1}(x - y) D_{2}(z - x) - D_{1}(x - y) D^{F}_{2}(z - x)
\nonumber \\
+ D^{(-)}_{1}(x - y) D^{(+)}_{2}(z - x) - D^{(+)}_{1}(x - y) D^{(-)}_{2}(z - x)
\nonumber \\
d^{(2)}_{D_{1},D_{2}}(x,y,z) \equiv 
- \bar{D}^{F}_{1}(x - y) D_{2}(y - z) + D_{1}(x - y) D^{F}_{2}(y - z)
\nonumber \\
+ D^{(+)}_{1}(x - y) D^{(-)}_{2}(y - z) - D^{(-)}_{1}(x - y) D^{(+)}_{2}(y - z)
\nonumber \\
d^{(3)}_{D_{1},D_{2}}(x,y,z) \equiv 
D^{F}_{1}(z - x) D_{2}(y - z) - D_{1}(z - x) D^{F}_{2}(y - z)
\nonumber \\
+ D^{(-)}_{1}(z - x) D^{(+)}_{2}(y - z) - D^{(+)}_{1}(z - x) D^{(-)}_{2}(y - z)
\label{d-tree}
\eea
where the dominant contribution corresponds to the choice
$
D_{1} = D_{2} = D_{m}.
$
As in the previous section we will use the alternative notation:
\be
d^{(j)}(D_{1},D_{2}) = d^{(j)}_{D_{1},D_{2}}.
\ee
The causal support properties follow from the alternative formulas
\bea
d^{(1)}_{D_{1},D_{2}}(x,y,z) =
D^{\rm ret}_{1}(x - y) D^{\rm ret}_{2}(z - x) 
- D^{\rm adv}_{1}(x - y) D^{\rm adv}_{2}(z - x)
\nonumber \\
d^{(2)}_{D_{1},D_{2}}(x,y,z) = D^{\rm ret}_{1}(y - x) D^{\rm ret}_{2}(z - y) 
- D^{\rm adv}_{1}(y - x) D^{\rm adv}_{2}(z - y)
\nonumber \\
d^{(3)}_{D_{1},D_{2}}(x,y,z) = D^{\rm ret}_{1}(z - x) D^{\rm ret}_{2}(y - z) 
- D^{\rm adv}_{1}(z - x) D^{\rm adv}_{2}(y - z).
\eea

The order of singularity of these distributions is
$
\omega = - 2
$.
We can define associated distributions if we replace
$
D_{1} \mapsto \partial_{\alpha}D_{1}
$,
etc. 
\bea
{\cal D}^{2}_{\alpha}d^{(1)}_{D_{1},D_{2}} =
d^{(1)}_{D_{1},\partial_{\alpha}D_{2}}, 
\qquad
{\cal D}^{3}_{\alpha}d^{(1)}_{D_{1},D_{2}} =
d^{(1)}_{\partial_{\alpha}D_{1},D_{2}}, 
\nonumber \\
{\cal D}^{1}_{\alpha}d^{(2)}_{D_{1},D_{2}} =
d^{(2)}_{D_{1},\partial_{\alpha}D_{2}}, 
\qquad
{\cal D}^{3}_{\alpha}d^{(2)}_{D_{1},D_{2}} =
d^{(2)}_{\partial_{\alpha}D_{1},D_{2}}, 
\nonumber \\
{\cal D}^{3}_{\alpha}d^{(3)}_{D_{1},D_{2}} =
d^{(3)}_{D_{1},\partial_{\alpha}D_{2}}, 
\qquad
{\cal D}^{2}_{\alpha}d^{(3)}_{D_{1},D_{2}} =
d^{(3)}_{\partial_{\alpha}D_{1},D_{2}}.
\eea
We have
\bea
{\partial \over \partial x^{\alpha}}d^{(1)} = 
( {\cal D}^{3}_{\alpha} - {\cal D}^{2}_{\alpha})d^{(1)},
\qquad
{\partial \over \partial y^{\alpha}}d^{(1)} =
- {\cal D}^{3}_{\alpha}d^{(1)}
\qquad
{\partial \over \partial z^{\alpha}}d^{(1)} =
{\cal D}^{2}_{\alpha}d^{(1)}
\nonumber \\
{\partial \over \partial x^{\alpha}}d^{(2)} = 
{\cal D}^{3}_{\alpha}d^{(2)},
\qquad
{\partial \over \partial y^{\alpha}}d^{(2)} =
({\cal D}^{1}_{\alpha} - {\cal D}^{3}_{\alpha})d^{(2)}
\qquad
{\partial \over \partial z^{\alpha}}d^{(2)} =
- {\cal D}^{1}_{\alpha}d^{(2)}
\nonumber \\
{\partial \over \partial x^{\alpha}}d^{(3)} = 
- {\cal D}^{2}_{\alpha}d^{(3)},
\qquad
{\partial \over \partial y^{\alpha}}d^{(3)} =
{\cal D}^{1}_{\alpha}d^{(3)}
\qquad
{\partial \over \partial z^{\alpha}}d^{(3)} =
({\cal D}^{2}_{\alpha} - {\cal D}^{1}_{\alpha})d^{(3)}.
\eea

The causal splitting of the distributions 
$
d^{(j)}
$
is elementary:
\bea
d^{(1)adv}_{D_{1},D_{2}}(x,y,z) = D^{\rm ret}_{1}(x - y) D^{\rm ret}_{2}(z - x),
\quad
d^{(1)ret}_{D_{1},D_{2}}(x,y,z) = D^{\rm adv}_{1}(x - y) D^{\rm adv}_{2}(z - x)
\nonumber \\
d^{(2)adv}_{D_{1},D_{2}}(x,y,z) = D^{\rm ret}_{1}(y - x) D^{\rm ret}_{2}(z - y),
\quad 
d^{(2)ret}_{D_{1},D_{2}}(x,y,z) =  D^{\rm adv}_{1}(y - x) D^{\rm adv}_{2}(z - y)
\nonumber \\
d^{(3)adv}_{D_{1},D_{2}}(x,y,z) = D^{\rm ret}_{1}(z - x) D^{\rm ret}_{2}(y - z),
\quad
d^{(3)ret}_{D_{1},D_{2}}(x,y,z) = D^{\rm adv}_{1}(z - x) D^{\rm adv}_{2}(y - z)
\label{split4}
\eea
and similar relations for the associated distributions
$
{\cal D}^{2}_{\alpha}d^{(1)}_{D_{1},D_{2}}
$,
etc. For the the Feynman propagators we have
\bea
d^{(1)F}_{D_{1},D_{2}}(x,y,z) =
D^{F}_{1}(x - y) D^{F}_{2}(z - x) 
\nonumber \\
d^{(2)F}_{D_{1},D_{2}}(x,y,z) = D^{F}_{1}(y - x) D^{F}_{2}(y - z)
\nonumber \\
d^{(3)}_{D_{1},D_{2}}(x,y,z) = D^{F}_{1}(z - x) D^{F}_{2}(y - z) 
\label{split5}
\eea
and it follows that these contributions do not produce anomalies. 

Another type of tree contributions comes from the last term of (\ref{T2ren}) i.e. the finite renormalizations.
We have the generic form
\be
N^{A,B}(x) = \sum:a_{1}(x) a_{2}(x) a_{3}(x) a_{4}(x):
\label{N-AB}
\ee
and when commuting with 
$
C(z) = :c_{1}(z)c_{2}(z)c_{3}(z):
$
we can have one, two or three contractions corresponding to tree, one-loop and two-loops contributions
\bea
T^{N}(A(x),B(y),C(z)) = T^{N}_{(0)}(A(x),B(y),C(z)) + T^{N}_{(1)} (A(x),B(y),C(z))
\nonumber\\
+ T^{N}_{(2)}(A(x),B(y),C(z)) + \cdots
\label{TN}
\eea
where $\cdots$ are the un-contracted terms. The relevant distributions for the tree contributions are 
\bea
d_{1}(x,y,z) = \delta(y - z)~D_{m}(x - y)
\nonumber \\
d_{2}(x,y,z) = \delta(z - x)~D_{m}(y - z)
\nonumber \\
d_{3}(x,y,z) = \delta(x - y)~D_{m}(y - z)
\label{dN}
\eea
where the dominant contributions correspond to the same positive mass. These contributions can produce anomalies by the same 
mechanism as for the tree contribution from the second order of perturbation theory.

%\newpage
\subsection{One-Loop Graphs: Triangle Type}
We consider again the tree contribution given by (\ref{T0}). When we commute this operator with
$
C(z) = :c_{1}(z)c_{2}(z)c_{3}(z):
$
we can contract a factor $a$ with one of the factors $c$ and one of the factors $b$ with another $c$; 
in terms of Feynman graphs it corresponds to triangle graphs.
We describe the relevant distributions with causal support. 

First, we take
$
D_{j} = D_{m_{j}}, j = 1,2,3
$
and define
\bea
d_{D_{1},D_{2},D_{3}}(x,y,z) \equiv \bar{D}^{F}_{3}(x - y) 
[ D^{(-)}_{2}(z - x) D^{(+)}_{1}(y - z) - D^{(+)}_{2}(z - x) D^{(-)}_{1}(y - z)
]
\nonumber \\
+ D^{F}_{1}(y - z) 
[ D^{(-)}_{3}(x - y) D^{(+)}_{2}(z - x) - D^{(+)}_{3}(x - y) D^{(-)}_{2}(z - x)
]
\nonumber \\
+ D^{F}_{2}(z - x) 
[ D^{(-)}_{1}(y - z) D^{(+)}_{3}(x - y) - D^{(+)}_{1}(y - z) D^{(-)}_{3}(x - y)
]
\eea
which also with causal support; indeed we have the alternative forms
\bea
d_{D_{1},D_{2},D_{3}}(x,y,z) = - D^{\rm ret}_{3}(x - y) 
[ D^{(-)}_{2}(z - x) D^{(+)}_{1}(y - z) - D^{(+)}_{2}(z - x) D^{(-)}_{1}(y - z)
]
\nonumber \\
+ D^{\rm adv}_{1}(y - z) 
[ D^{(-)}_{3}(x - y) D^{(+)}_{2}(z - x) - D^{(+)}_{3}(x - y) D^{(-)}_{2}(z - x)
]
\nonumber \\
+ D^{\rm adv}_{2}(z - x) 
[ D^{(-)}_{1}(y - z) D^{(+)}_{3}(x - y) - D^{(+)}_{1}(y - z) D^{(-)}_{3}(x - y)
]
\eea
and
\bea
d_{D_{1},D_{2},D_{3}}(x,y,z) = - D^{\rm adv}_{3}(x - y) 
[ D^{(-)}_{2}(z - x) D^{(+)}_{1}(y - z) - D^{(+)}_{2}(z - x) D^{(-)}_{1}(y - z)
]
\nonumber \\
+ D^{\rm ret}_{1}(y - z) 
[ D^{(-)}_{3}(x - y) D^{(+)}_{2}(z - x) - D^{(+)}_{3}(x - y) D^{(-)}_{2}(z - x)
]
\nonumber \\
+ D^{\rm ret}_{2}(z - x) 
[ D^{(-)}_{1}(y - z) D^{(+)}_{3}(x - y) - D^{(+)}_{1}(y - z) D^{(-)}_{3}(x - y)
]
\eea
from which it follows that the distribution
$
d_{D_{1},D_{2},D_{3}}(x,y,z)
$
is null outside the causal cone
$
\{ (x,y,z) | x - z \in V^{+}, y - z \in V^{+}  \} \cup 
\{ (x,y,z) | x - z \in V^{-}, y - z \in V^{-}  \}
$. 
These distributions have the singularity order
$
\omega(d_{D_{1},D_{2},D_{3}}) = - 2
$.

As in the previous Section we use the alternative notation
\be
d_{123} \equiv  d(D_{1},D_{2},D_{3}) \equiv d_{D_{1},D_{2},D_{3}} 
\label{d123}
\ee
and when there is no ambiguity about the distributions
$
D_{j}
$
we simply denote
$
d = d_{123}
$.
There are some associated distributions obtained from
$
d_{D_{1},D_{2},D_{3}}(x,y,z)
$
applying derivatives on the factors
$
D_{j} = D_{m_{j}}, j = 1,2,3
$. 
We also denote
\bea
{\cal D}_{1}^{\mu}d_{D_{1},D_{2},D_{3}} \equiv
d_{\partial^{\mu}D_{1},D_{2},D_{3}},\quad
%\nonumber \\
{\cal D}_{2}^{\mu}d_{D_{1},D_{2},D_{3}} \equiv
d_{D_{1},\partial^{\mu}D_{2},D_{3}},\quad
%\nonumber \\
{\cal D}_{3}^{\mu}d_{D_{1},D_{2},D_{3}} \equiv
d_{D_{1},D_{2},\partial^{\mu}D_{3}},
\eea
and so on for more derivatives
$
\partial_{\alpha}
$
distributed in an arbitrary way on the factors
$
D_{j} = D_{m_{j}}, j = 1,2,3
$.
We note that we have:
\bea
{\partial \over \partial x_{\mu}}d = 
( {\cal D}_{3}^{\mu} - {\cal D}_{2}^{\mu})d, \quad
%\nonumber \\
{\partial \over \partial y_{\mu}}d =
( {\cal D}_{1}^{\mu} - {\cal D}_{3}^{\mu})d, \quad
%\nonumber \\
{\partial \over \partial z_{\mu}}d =
( {\cal D}_{2}^{\mu} - {\cal D}_{1}^{\mu})d.
\label{d123-p}
\eea

It is known that these distributions can be causally split in such a way that
the order of singularity, translation invariance and Lorentz covariance are
preserved. The same will be true for the corresponding Feynman distributions.
Because 
$
\omega(d_{123}) = - 2
$
and
$
\omega({\cal D}_{i}^{\mu}d_{123}) = - 1
$
the corresponding advanced, retarded and Feynman distributions are unique. For
more derivatives we have some freedom of redefinition.

As in the previous Section, let us consider the case
$
D_{1} = D_{2} = D_{3} = D_{m},~m > 0
$
and study the corresponding distribution
$
d_{m,m,m}.
$
We consider it as distribution in two variables
$
X \equiv x - z,\quad Y \equiv y - z
$
and we will need its Fourier transform. The computation is essentially done in
\cite{Sc1} and gives the following formula:
\be
\tilde{d}_{m,m,m}(p,q) = {1\over 8 (2\pi)^{5}} {1 \over \sqrt{N}}~
[\epsilon(p_{0}) \theta(p^{2} - 4 m^{2})~ln_{1}
+ \epsilon(q_{0}) \theta(q^{2} - 4 m^{2})~ln_{2} 
+ \epsilon(P_{0}) \theta(P^{2} - 4 m^{2})~ln_{3} ]
\label{d-mmm}
\ee
where
\bea
ln_{1} \equiv ln\left({P\cdot q + \sqrt{N (1 - 4 m^{2}/p^{2})} \over
P\cdot q - \sqrt{N (1 - 4 m^{2}/p^{2})}}\right)
\nonumber \\
ln_{2} \equiv ln\left({P\cdot p + \sqrt{N (1 - 4 m^{2}/q^{2})} \over
P\cdot p - \sqrt{N (1 - 4 m^{2}/q^{2})}}\right)
\nonumber \\
ln_{3} \equiv ln\left({- p\cdot q + \sqrt{N (1 - 4 m^{2}/P^{2})} \over
- p\cdot q - \sqrt{N (1 - 4 m^{2}/P^{2})}}\right)
\eea
with the notations
$
P = p + q
$
and
$
N \equiv (p\cdot q)^{2} - p^{2} q^{2}.
$
%\newpage
We give here and example of the use of such a causal distribution. By direct computation we can prove
\newpage
\begin{thm}
In the Yang-Mills sector the dominant contribution (i.e. of maximal order of singularity) for one-loop triangle graphs is:
\bea
D_{YM}^{[\mu],[\nu],[\rho]}(x,y,z)_{(1)} = i
[ {\cal D}_{1}^{\mu}{\cal D}_{1}^{\nu}{\cal D}_{2}^{\rho}
+ {\cal D}_{1}^{\rho}{\cal D}_{2}^{\mu}{\cal D}_{2}^{\nu}
+ {\cal D}_{1}^{\mu}{\cal D}_{1}^{\rho}{\cal D}_{3}^{\nu}
\nonumber\\
+ {\cal D}_{1}^{\nu}{\cal D}_{3}^{\mu}{\cal D}_{3}^{\rho}
+ {\cal D}_{2}^{\nu}{\cal D}_{2}^{\rho}{\cal D}_{3}^{\mu}
+ {\cal D}_{2}^{\mu}{\cal D}_{3}^{\nu}{\cal D}_{3}^{\rho}
\nonumber \\
+ {\cal D}_{1}^{\nu}{\cal D}_{1}^{\rho}{\cal D}_{2}^{\mu}
+ {\cal D}_{1}^{\nu}{\cal D}_{2}^{\mu}{\cal D}_{2}^{\rho}
+ {\cal D}_{1}^{\nu}{\cal D}_{1}^{\rho}{\cal D}_{3}^{\mu}
\nonumber\\
+ {\cal D}_{1}^{\rho}{\cal D}_{3}^{\mu}{\cal D}_{3}^{\nu}
+ {\cal D}_{2}^{\mu}{\cal D}_{2}^{\rho}{\cal D}_{3}^{\nu}
+ {\cal D}_{2}^{\rho}{\cal D}_{3}^{\mu}{\cal D}_{3}^{\nu}
\nonumber \\
- 2 {\cal D}_{1}^{\mu}{\cal D}_{2}^{\nu}{\cal D}_{3}^{\rho}
\nonumber \\
+ 2 ({\cal D}_{1}^{\mu}{\cal D}_{2}^{\rho}{\cal D}_{3}^{\nu}
+ {\cal D}_{1}^{\rho}{\cal D}_{2}^{\nu}{\cal D}_{3}^{\mu}
+ {\cal D}_{1}^{\nu}{\cal D}_{2}^{\mu}{\cal D}_{3}^{\rho})
\nonumber \\
+ \eta^{\mu\nu} {\cal D}_{1}^{\rho}{\cal D}_{1}\cdot{\cal D}_{2}
+ \eta^{\mu\nu} {\cal D}_{2}^{\rho}{\cal D}_{1}\cdot{\cal D}_{2}
+ \eta^{\mu\rho} {\cal D}_{1}^{\nu}{\cal D}_{1}\cdot{\cal D}_{3}
\nonumber \\
+ \eta^{\mu\rho} {\cal D}_{3}^{\nu}{\cal D}_{1}\cdot{\cal D}_{3}
+ \eta^{\nu\rho} {\cal D}_{2}^{\mu}{\cal D}_{2}\cdot{\cal D}_{3}
+ \eta^{\nu\rho} {\cal D}_{3}^{\mu}{\cal D}_{2}\cdot{\cal D}_{3}
\nonumber \\
- (\eta^{\mu\nu} {\cal D}_{3}^{\rho}{\cal D}_{1}\cdot{\cal D}_{3}
+ \eta^{\mu\nu} {\cal D}_{3}^{\rho}{\cal D}_{2}\cdot{\cal D}_{3}
+ \eta^{\mu\rho} {\cal D}_{2}^{\nu}{\cal D}_{1}\cdot{\cal D}_{2}
\nonumber \\
+ \eta^{\mu\rho} {\cal D}_{2}^{\nu}{\cal D}_{2}\cdot{\cal D}_{3}
+ \eta^{\nu\rho} {\cal D}_{1}^{\mu}{\cal D}_{1}\cdot{\cal D}_{2}
+ \eta^{\nu\rho} {\cal D}_{1}^{\mu}{\cal D}_{1}\cdot{\cal D}_{3}) ]d(x,y,z)
\nonumber\\
f^{(0)}_{abc}~u_{a}(x)~u_{b}(y)~u_{c}(z)
\label{D3}
\eea
where
\be
f^{(0)}_{[abc]} = f_{eap} f_{ebq} f_{cpq}.
\ee
\end{thm}
%\newpage
\subsection{One-Loop Graphs: One-Particle Reducible Type}

Such contributions have two sources: 
(a) from the one-loop contribution of (\ref{T2ren}) of the generic form
\be
T(A(x),B(y))_{(1)} = \sum p_{j}(\partial)d_{2}^{F}(x - y)~:a(x)b(y):
\label{T1}
\ee
Commuting with
$
C(z) = :c_{1}(z)c_{2}(z)c_{3}(z):
$
we contract the factor $a$ (or the factor $b$) with one of the factors $c$;
(b) from (\ref{T0}) commuting with 
$
C(z) = :c_{1}(z)c_{2}(z)c_{3}(z):
$
we contract two factors $a$ (or two factors $b$) with the factors $c$. 
These relevant causal distributions are of the type (\ref{d-tree}) namely
\be
d^{(j)}_{m,m,m} = d_{D_{m},d_{m,m}}^{(j)} = d^{(j)}(D_{m},d_{m,m}),\qquad 
f^{(j)}_{m,m,m} = d_{d_{m,m},D_{m}}^{(j)} = d^{(j)}(d_{m,m},D_{m}),~j = 1,2,3
\label{df}
\ee
where 
$
d_{m,m}
$
is defined by (\ref{d2}) for equal masses
$
m_{1} = m_{2} = m.
$
%\newpage
We illustrate the use of these distributions by the following
\begin{thm}
In the Yang-Mills sector the dominant contribution (i.e. of maximal order of singularity) for one-loop, one-particle reducible graphs is
\bea
D(T^{\mu}(x),T^{\nu}(y);T^{\rho}(z))_{1 PR} = - {i\over 3}~(f^{(0)}_{abc} + f^{(3)}_{abc} + f^{(4)}_{abc})~
\nonumber\\
\Bigl[ 
 {\cal D}^{\mu}_{2}( {\cal D}^{\nu}_{1}{\cal D}^{\rho}_{1} - \eta^{\nu\rho} {\cal D}_{1}^{2})d^{(3)}(x,y,z)~u_{a}(x)u_{b}(x)u_{c}(y)
\nonumber\\
+ {\cal D}^{\nu}_{1}( {\cal D}^{\mu}_{2}{\cal D}^{\rho}_{2} - \eta^{\mu\rho} {\cal D}_{2}^{2})f^{(3)}(x,y,z)~u_{a}(x)u_{b}(y)u_{c}(y)
\nonumber\\
+ {\cal D}^{\mu}_{3}( {\cal D}^{\nu}_{1}{\cal D}^{\rho}_{1} - \eta^{\nu\rho} {\cal D}_{1}^{2})d^{(2)}(x,y,z)~u_{a}(x)u_{b}(x)u_{c}(z)
\nonumber\\
+ {\cal D}^{\rho}_{1}( {\cal D}^{\mu}_{3}{\cal D}^{\nu}_{3} - \eta^{\mu\nu} {\cal D}_{3}^{2})f^{(2)}(x,y,z)~u_{a}(x)u_{b}(z)u_{c}(z)
\nonumber\\
+ {\cal D}^{\nu}_{3}( {\cal D}^{\mu}_{2}{\cal D}^{\rho}_{2} - \eta^{\mu\rho} {\cal D}_{2}^{2})d^{(1)}(x,y,z)~u_{y}(x)u_{b}(y)u_{c}(z)
\nonumber\\
+ {\cal D}^{\rho}_{2}( {\cal D}^{\mu}_{3}{\cal D}^{\nu}_{3} - \eta^{\mu\nu} {\cal D}_{3}^{2})f^{(1)}(x,y,z)~u_{a}(y)u_{b}(z)u_{c}(z)\Bigl]
\eea
where 
$
f^{(0)}_{abc}
$
has been defined in the previous theorem and 
\be
f^{(3)}_{[abc]} = f^{\prime}_{epa} f^{\prime}_{eqb} f^{\prime}_{pqc}, \qquad
f^{(4)}_{[abc]} = i~Tr([ t_{a}^{\epsilon},t_{b}^{\epsilon}]t_{c}^{\epsilon}) =
f_{abd}~g^{(2)}_{cd}.
\ee
\end{thm}

We also have loop contributions of one-particle reducible type associated to the finite renormalizations (the last term) from 
(\ref{T2ren}). We commute an expression of the type (\ref{N-AB}) with 
$
C(z) = :c_{1}(z)c_{2}(z)c_{3}(z):
$
and take two contractions and obtain
$
T^{N}_{(1)}(A(x),B(y),C(z)).
$ 

The relevant causal distributions are 
\bea
f_{1}(x,y,z) = \delta(y - z)~d_{m,m}(x - y)
\nonumber \\
f_{2}(x,y,z) = \delta(z - x)~d_{m,m}(y - z)
\nonumber \\
f_{3}(x,y,z) = \delta(x - y)~d_{m,m}(y - z)
\eea
with 
\be
\omega(f_{j}) = 0.
\label{deg-f}
\ee
We consider them (as before) as distributions in two variables
$
X \equiv x - z,~Y \equiv y - z
$
and the Fourier transforms are:
\be
\tilde{f}_{1}(p,q) = {1 \over (2\pi)^{2}}~\tilde{d}_{m,m}(p),\quad
\tilde{f}_{2}(p,q) = {1 \over (2\pi)^{2}}~\tilde{d}_{m,m}(q),\quad
\tilde{f}_{3}(p,q) = {1 \over (2\pi)^{2}}~\tilde{d}_{m,m}(P)
\label{f}
\ee
but these contributions do not produce anomalies. The same is true for the last contribution in (\ref{TN}).
We will need in the next Section the distributions:
\bea
f_{1}^{\prime}(x,y,z) = \delta(y - z)~d_{m,m}^{\prime}(x - y)
\nonumber \\
f_{2}^{\prime}(x,y,z) = \delta(z - x)~d_{m,m}^{\prime}(y - z)
\nonumber \\
f_{3}^{\prime}(x,y,z) = \delta(x - y)~d_{m,m}^{\prime}(y - z).
\eea
\subsection{Two-Loop Graphs}
The associated causal distributions are
$
d(d_{m,m},D_{m},D_{m}), d(D_{m},d_{m,m},D_{m}), d(D_{m},D_{m},d_{m,m})
$
in the notation (\ref{d123}).
\newpage
\section{Causal Splitting in the Third Order for Triangle Contributions}

We denote for simplicity
\bea
d_{i}^{\mu} \equiv {\cal D}_{i}^{\mu} d, 
\nonumber\\
d_{ij}^{\mu\nu} \equiv {\cal D}_{i}^{\mu} {\cal D}_{j}^{\nu}d, \qquad d_{ij} \equiv \eta_{\mu\nu}~d_{ij}^{\mu\nu}
\nonumber\\
d_{ijk}^{\mu\nu\rho} \equiv {\cal D}_{i}^{\mu} {\cal D}_{j}^{\nu} {\cal D}_{k}^{\rho} d, 
\qquad d_{ijk}^{\mu} \equiv \eta_{\nu\rho}  d_{ijk}^{\mu\nu\rho}
%\nonumber\\
%d_{ijkl}^{\mu\nu\rho\sigma} \equiv {\cal D}_{i}^{\mu} {\cal D}_{j}^{\nu} {\cal D}_{k}^{\rho} {\cal D}_{l}^{\sigma}d,
%\qquad d_{ijkl}^{\mu\nu} \equiv \eta_{\rho\sigma} d_{ijkl}^{\mu\nu\rho\sigma}
\label{dijk}
\eea
and we have the following orders of singularity:
\be
\omega(d_{j}^{\mu}) = - 1, \quad \omega(d_{jk}^{\mu\nu}) = 0, \quad \omega(d_{jkl}^{\mu\nu\rho}) = 1.
\label{deg}
\ee

To perform the computation we need an explicit formula for the Fourier transform of these distributions. We remind the analysis 
from \cite{loop}. From Lorentz covariance considerations the Fourier transform should be of the
form:
\be
\tilde{d}^{\mu}_{j}(p,q) = - i~[p^{\mu}~\tilde{A}_{j}(p,q) +
q^{\mu}~\tilde{B}_{j}(p,q)]
\label{d11}
\ee
where the scalar functions
$
\tilde{A}_{j}
$
and
$
\tilde{B}_{j}
$
depend in fact only on the Lorentz invariants:
$
p^{2}, q^{2}, p\cdot q.
$
It is not hard to obtain the explicit formulas
\bea
\tilde{A}_{3}(p,q) = - {q^{2} p\cdot P\over 2 N} \tilde{d}_{m,m,m}(p,q) 
+ {q^{2}\over N} [ \tilde{f}_{3}(p,q) - \tilde{f}_{2}(p,q)] 
+ {p\cdot q \over N} [ \tilde{f}_{3}(p,q) - \tilde{f}_{1}(p,q)] 
\nonumber \\
\tilde{B}_{3}(p,q) = - \tilde{A}_{3}(q,p)
\label{d12}
\eea

The expression
$
\tilde{d}^{\mu}_{2}(p,q)
$
can be obtained from the preceding expression
$
\tilde{d}^{\mu}_{3}(p,q)
$
applying the transformation
\be
p \rightarrow - p,~q \rightarrow P
\label{T23}
\ee
and expression
$
\tilde{d}^{\mu}_{1}(p,q)
$
can be obtained from the expression
$
\tilde{d}^{\mu}_{2}(p,q)
$
applying the transformation
\be
p \rightarrow - q,~q \rightarrow - p.
\label{T12}
\ee

Now we have the following generic form of the Fourier transform:
\be
\tilde{d}^{\mu\nu}_{jk}(p,q) = - [p^{\mu}p^{\nu}~\tilde{A}_{jk}(p,q) +
q^{\mu}q^{\nu}~\tilde{B}_{jk}(p,q) + p^{\mu}q^{\nu}~\tilde{C}^{(1)}_{jk}(p,q)
+ q^{\mu}p^{\nu}~\tilde{C}^{(2)}_{jk}(p,q)] + \eta^{\mu\nu}~\tilde{D}_{jk}(p,q)
\label{d21}
\ee
where, as before, the scalar functions
$
A, B, C, D
$
depend only on the Lorentz invariants. 

It is a long but straightforward computation to derive the following
expressions:
\bea
\tilde{A}_{33}(p,q) = {3 q^{2} \over 2 N^{2}} \alpha(p,q) 
+ {1 \over N} \alpha_{2}(p,q)
- {q^{2}\over N} \tilde{f}_{3}(p,q) 
+  {m^{2} q^{2} \over 2N} \tilde{d}_{m,m,m}(p,q)
\nonumber \\
\tilde{B}_{33}(p,q) = {3 p^{2} \over 2 N^{2}} \alpha(p,q) 
+ {1 \over N} \alpha_{1}(p,q)
- {p^{2}\over N} \tilde{f}_{3}(p,q) 
+  {m^{2} p^{2} \over 2N} \tilde{d}_{m,m,m}(p,q) = \tilde{A}_{33}(q,p)
\nonumber \\
\tilde{C}_{33}^{(1)}(p,q) = \tilde{C}_{33}^{(2)}(p,q) 
= - {3 p\cdot q \over 2 N^{2}} \alpha(p,q) 
- {1 \over N} \alpha_{3}(p,q)
+ {p \cdot q\over N} \tilde{f}_{3}(p,q) 
-  {m^{2} p \cdot q \over 2N} \tilde{d}_{m,m,m}(p,q)
\label{d22}
\eea
where
\bea
\alpha_{1}(p,q) = {1\over 4}~(p^{2})^{2}~\tilde{d}_{m,m,m}(p,q)
+ {1 \over 2}~(p^{2} - p \cdot q)~\tilde{f}_{2}(p,q)
- \left(p^{2} - {1\over 2}~p \cdot q\right)~\tilde{f}_{3}(p,q)
\nonumber \\
\alpha_{2}(p,q) = {1\over 4}~(q^{2})^{2}~\tilde{d}_{m,m,m}(p,q)
+ {1 \over 2}~(q^{2} - p \cdot q)~\tilde{f}_{1}(p,q)
- \left(q^{2} - {1\over 2}~p \cdot q\right)~\tilde{f}_{3}(p,q)
\nonumber \\
\alpha_{3}(p,q) = - {1\over 4}~p^{2} q^{2}~\tilde{d}_{m,m,m}(p,q)
- {1 \over 2}~p^{2}~\tilde{f}_{1}(p,q)- {1 \over 2}~q^{2}~\tilde{f}_{2}(p,q) 
\nonumber \\
+ {1\over 2} (p^{2} + q^{2} - p \cdot q)~\tilde{f}_{3}(p,q)
\label{23}
\eea
and
\be
\alpha(p,q) = q^{2}~\alpha_{1}(p,q) + p^{2}~\alpha_{2}(p,q) 
- 2 p \cdot q~\alpha_{3}(p,q).
\label{24}
\ee

The expression
$
\tilde{d}^{\mu}_{22}(p,q)
$
can be obtained from the preceding expression
$
\tilde{d}^{\mu}_{33}(p,q)
$
applying the transformation (\ref{T23}) and expression
$
\tilde{d}^{\mu}_{11}(p,q)
$
can be obtained from the expression
$
\tilde{d}^{\mu}_{22}(p,q)
$
applying the transformation (\ref{T12}). 

In the same way we have
\be
\tilde{D}_{12}(p,q) = - {1 \over 2 N} [q^{2}  \beta_{1}(p,q)
+ p^{2} \beta_{2}(p,q) ] + {p \cdot q \over 2 N} [\beta_{3}(p,q)
+ \beta_{4}(p,q) ] - {1\over 2} \beta_{5}(p,q)
\label{d31}
\ee
and
\bea
\tilde{A}_{12}(p,q) = - {1 \over N} [3 q^{2}  \tilde{D}_{12}(p,q)
+ q^{2} \beta_{5}(p,q) - \beta_{2}(p,q) ]
\nonumber \\
\tilde{B}_{12}(p,q) = - {1 \over N} [3 p^{2}  \tilde{D}_{12}(p,q)
+ p^{2} \beta_{5}(p,q) - \beta_{1}(p,q) ]
\nonumber \\
\tilde{C}_{12}^{(1)}(p,q) = {1 \over N} [3 p \cdot q  \tilde{D}_{12}(p,q)
- \beta_{3}(p,q) + p \cdot q \beta_{5}(p,q) ]
\nonumber \\
\tilde{C}_{12}^{(2)}(p,q) = {1 \over N} [3 p \cdot q  \tilde{D}_{12}(p,q)
- \beta_{4}(p,q) + p \cdot q \beta_{5}(p,q) ].
\label{d32}
\eea

Here we have the notations:
\bea
\beta_{1}(p,q) = - {1\over 4}~p^{2}~(p^{2} + 2 p \cdot q)~\tilde{d}_{m,m,m}(p,q)
- {1\over 2} (p^{2} - p \cdot q)~\tilde{f}_{2}(p,q) 
- {1\over 2} (p\cdot q)~\tilde{f}_{3}(p,q)
\nonumber \\
\beta_{2}(p,q) = - {1\over 4}~q^{2}~(q^{2} + 2 p \cdot q)~\tilde{d}_{m,m,m}(p,q)
- {1\over 2} (q^{2} - p \cdot q)~\tilde{f}_{1}(p,q) 
- {1\over 2} (p\cdot q)~\tilde{f}_{3}(p,q)
\nonumber \\ 
\beta_{3}(p,q) = 
- {1\over 4}~(p^{2} + 2 p \cdot q)~(q^{2} + 2 p \cdot q)~\tilde{d}_{m,m,m}(p,q)
\nonumber \\
- {1\over 2} (p^{2} + 2 p \cdot q)~\tilde{f}_{1}(p,q) 
- {1\over 2} (q^{2} + 2 p \cdot q)~\tilde{f}_{2}(p,q) 
+ {1\over 2} (p^{2} + q^{2} + 3 p\cdot q)~\tilde{f}_{3}(p,q)
\nonumber \\
\beta_{4}(p,q) = 
- {1\over 4}~p^{2}~q^{2}~\tilde{d}_{m,m,m}(p,q)
+ {1\over 2} p^{2}~\tilde{f}_{1}(p,q) 
+ {1\over 2} q^{2}~\tilde{f}_{2}(p,q) 
- {1\over 2} (p^{2} + q^{2} + p\cdot q)~\tilde{f}_{3}(p,q)
\nonumber \\
\beta_{5}(p,q) = 
- {1\over 2}~(p + q)^{2}~\tilde{d}_{m,m,m}(p,q)
- \tilde{f}_{1}(p,q) - \tilde{f}_{2}(p,q) 
+ m^{2}~\tilde{d}_{m,m,m}(p,q)
\nonumber \\
\label{d33}
\eea
The expression
$
\tilde{d}^{\mu}_{13}(p,q)
$
can be obtained from the preceding expression
$
\tilde{d}^{\mu}_{12}(p,q)
$
applying the transformation (\ref{T23}) and expression
$
\tilde{d}^{\mu}_{23}(p,q)
$
can be obtained from the expression
$
\tilde{d}^{\mu}_{13}(p,q)
$
applying the transformation (\ref{T12}).

Using these formulas we will be able to perform the central causal splitting.

We start with the simplest case. 
\begin{thm}
The following relations are true
\be
{\partial \over \partial x^{\rho}} {\cal D}_{1}^{\rho}d = {\cal D}_{1} \cdot {\cal D}_{3}d - {\cal D}_{1} \cdot {\cal D}_{2}d
\label{1.1}
\ee
\be
{\partial \over \partial y^{\rho}} {\cal D}_{1}^{\rho}d = -  {\cal D}_{1} \cdot {\cal D}_{3}d - m^{2}~d + 2~f_{1},
\label{1.2}
\ee
\be
{\partial \over \partial z^{\rho}} {\cal D}_{1}^{\rho}d = {\cal D}_{1} \cdot {\cal D}_{2}d + m^{2}~d - 2~f_{1}
\label{1.3}
\ee
and another two sets of relations which can be obtained by circular permutations.

After the central causal splitting we obtain:
\be
{\partial \over \partial x^{\rho}} ({\cal D}_{1}^{\rho}d)^{F} = ({\cal D}_{1} \cdot {\cal D}_{3}d)^{F} 
- ({\cal D}_{1} \cdot {\cal D}_{2}d)^{F}
\label{1.1F}
\ee
\be
{\partial \over \partial y^{\rho}} ({\cal D}_{1}^{\rho}d)^{F} = -  ({\cal D}_{1} \cdot {\cal D}_{3}d)^{F} - m^{2}~d^{F} + 2~f^{F}_{1} 
+ A~\delta,
\label{1.2F}
\ee
\be
{\partial \over \partial z^{\rho}} ({\cal D}_{1}^{\rho}d)^{F} = ({\cal D}_{1} \cdot {\cal D}_{2}d)^{F} + m^{2}~d^{F} - 2~f^{F}_{1}
- A~\delta
\label{1.3F}
\ee
and another two sets of relations which can be obtained by circular permutations; here
$
\delta = \delta(x,y,z) = \delta(x - z) \delta(y - z)
$
and
$
A = {i \over 8 (2\pi)^{2}}.
$
\label{split1}
\end{thm}
{\bf Proof:} We illustrate the idea using the relation (\ref{1.2}); after we perform a Fourier transform:
\be
- i~q_{\mu}\tilde{d}^{\mu}_{1} = - \tilde{d}_{13} - m^{2} \tilde{d} + 2 \tilde{f}_{1}.
\ee
Because of (\ref{deg}) and (\ref{deg-f}) we must causally split
$
d
$
and
$
d_{1}^{\mu}
$
with formula (\ref{central1}) and
$
d_{13}
$
and
$
f_{1}
$
with formula (\ref{central2}). It follows that the anomaly
\be
\tilde{A}_{1} \equiv - i~q_{\rho}\tilde{a}^{\rho}_{1} + \tilde{a}_{13} + m^{2} \tilde{a} - 2 \tilde{f}^{adv}_{1}
\ee
is given by
\be
\tilde{A}_{1} = - {i m^{2} \over 2 \pi}~\int {dt \over t} \tilde{d}(tp, tq) = {i \over 8 (2\pi)^{6}}
\ee
and this gives (\ref{1.2F}). All other relations are causally split in the same way.
$\qed$
%\newpage

Next, we have a more complicated case. 
\begin{thm}
The following relations are true
\be
{\partial \over \partial x^{\rho}} {\cal D}_{i}^{\mu}{\cal D}_{1}^{\rho}d 
= {\cal D}_{i}^{\mu}{\cal D}_{1} \cdot {\cal D}_{3}d - {\cal D}_{i}^{\mu}{\cal D}_{1} \cdot {\cal D}_{2}d
\label{2.1}
\ee
\be
{\partial \over \partial y^{\rho}} {\cal D}_{i}^{\mu}{\cal D}_{1}^{\rho}d = -  {\cal D}_{i}^{\mu}{\cal D}_{1} \cdot {\cal D}_{3}d 
- m^{2}~{\cal D}_{i}^{\mu}d + f^{\mu}_{i1},
\label{2.2}
\ee
\be
{\partial \over \partial z^{\rho}} {\cal D}_{i}^{\mu}{\cal D}_{1}^{\rho}d = {\cal D}_{i}^{\mu}{\cal D}_{1} \cdot {\cal D}_{2}d 
+ m^{2}~{\cal D}_{i}^{\mu}d - f^{\mu}_{i1}
\label{2.3}
\ee
and another two sets of relations which can be obtained by circular permutations. Here
\be
f^{\mu}_{11} = (\partial^{\mu}_{1} + 2 \partial^{\mu}_{2})f_{1},\quad
f^{\mu}_{21} = - \partial^{\mu}_{1}f_{1},\quad
f^{\mu}_{31} = \partial^{\mu}_{1}f_{1}
\ee
and the rest by circular permutations. 

After central causal splitting we obtain:
\be
{\partial \over \partial x^{\rho}} ({\cal D}_{i}^{\mu}{\cal D}_{1}^{\rho}d)^{F} 
= ({\cal D}_{i}^{\mu}{\cal D}_{1} \cdot {\cal D}_{3}d)^{F} - ({\cal D}_{i}^{\mu}{\cal D}_{1} \cdot {\cal D}_{2}d)^{F}
\label{2.1F}
\ee
\be
{\partial \over \partial y^{\rho}} ({\cal D}_{i}^{\mu}{\cal D}_{1}^{\rho}d)^{F} = - ({\cal D}_{i}^{\mu}{\cal D}_{1} \cdot {\cal D}_{3}d)^{F} 
- m^{2}~({\cal D}_{i}^{\mu}d)^{F} + f^{\mu,F}_{i1} + A_{1}^{\mu},
\label{2.2F}
\ee
\be
{\partial \over \partial z^{\rho}} ({\cal D}_{i}^{\mu}{\cal D}_{1}^{\rho}d)^{F} = ({\cal D}_{i}^{\mu}{\cal D}_{1} \cdot {\cal D}_{2}d)^{F} 
+ m^{2}~({\cal D}_{i}^{\mu}d)^{F} - f^{\mu,F}_{i1} - A_{1}^{\mu}
\label{2.3F}
\ee
and another two sets of relations which can be obtained by circular permutations. Here
\be
A^{\mu}_{1} = B (\partial_{2}^{\mu} - \partial_{3}^{\mu})~\delta = B (\partial_{1}^{\mu} + 2 \partial_{2}^{\mu})~\delta
\ee
and the rest by circular permutations. We have defined
$
B \equiv {1 \over 3}~A.
$
\label{split2}
\end{thm}
{\bf Proof:}
We consider (\ref{2.2}): after the Fourier transform, we end up, as before, with the anomaly
\be
\tilde{A}_{j1} = - i~q_{\rho}\tilde{a}^{\mu\rho}_{j1} + \tilde{a}^{\mu}_{j13} + m^{2} \tilde{a}^{\mu}_{j} - \tilde{f}^{\mu,\rm adv}_{j1}
\ee
After the causal splitting we find out that
\be
\tilde{A}_{j1} = - {i m^{2} \over 2 \pi}~\int {dt \over t^{2}}(1 + t) \tilde{d}^{\mu}_{j}(tp, tq)
\ee
dependents only on $j$. We must use the formula (\ref{d11}) and we obtain:
\be
\tilde{\cal A}_{j}^{\mu}(p,q) = - { m^{2} \over 2\pi} 
\left[ p^{\mu}~\int {dt\over t}~\tilde{A}_{j}(tp, tq)
+ q^{\mu}~\int {dt\over t}~\tilde{B}_{j}(tp, tq)\right].
\ee
To compute the two integrals above we must use the formulas (\ref{d12}). For
instance we have:
\bea
\int {dt\over t}~\tilde{A}_{3}(tp, tq)
= - {q^{2} p\cdot P\over 2 N} \int {dt\over t}\tilde{d}_{m,m,m}(tp,tq) 
\nonumber \\
+ {q^{2}\over N} \int {dt\over t^{3}}
[ \tilde{f}_{3}(tp,tq) - \tilde{f}_{2}(tp,tq)] 
+ {p\cdot q \over N} \int {dt\over t^{3}}
[ \tilde{f}_{3}(tp,tq) - \tilde{f}_{1}(tp,tq)]. 
\eea
The first integral has been already computed at the preceding theorem. If we
use the expressions (\ref{f}) then we get
\be
\int {dt\over t^{3}}~\tilde{f}_{1}(tp,tq) = b(p^{2}),\qquad
\int {dt\over t^{3}}~\tilde{f}_{2}(tp,tq) = b(q^{2}),\qquad
\int {dt\over t^{3}}~\tilde{f}_{3}(tp,tq) = b(P^{2})
\ee
where
\be
b(k) \equiv {1\over (2\pi)^{2}}~\int {dt\over t^{3}}~\tilde{d}_{m,m}(tk).
\ee
The preceding integral can be computed using the explicit expression
(\ref{d-mm}) and the result is
\be
b(k) = b~k^{2},\qquad b \equiv - {1 \over 48 (2\pi)^{5} m^{2}}.
\label{b}
\ee
so after some simple substitutions we obtain the formulas from the statement.
$\qed$
%\newpage

Finally we have:
\begin{thm}
The following relations are true
\be
{\partial \over \partial x^{\rho}} {\cal D}_{i}^{\mu}{\cal D}_{j}^{\nu}{\cal D}_{1}^{\rho}d 
= {\cal D}_{i}^{\mu}{\cal D}_{j}^{\nu}{\cal D}_{1} \cdot {\cal D}_{3}d - {\cal D}_{i}^{\mu}{\cal D}_{j}^{\nu}{\cal D}_{1} \cdot {\cal D}_{2}d
\label{3.1}
\ee
\be
{\partial \over \partial y^{\rho}} {\cal D}_{i}^{\mu}{\cal D}_{j}^{\nu}{\cal D}_{1}^{\rho}d = 
-  {\cal D}_{i}^{\mu}{\cal D}_{j}^{\nu}{\cal D}_{1} \cdot {\cal D}_{3}d 
- m^{2}~{\cal D}_{i}^{\mu}{\cal D}_{j}^{\nu}d + f^{\mu\nu}_{ij1} - {2 m^{2} \over 3} C^{\mu\nu}_{1}~f^{\prime}_{1},
\label{3.2}
\ee
\be
{\partial \over \partial z^{\rho}} {\cal D}_{i}^{\mu}{\cal D}_{j}^{\nu}{\cal D}_{1}^{\rho}d = 
{\cal D}_{i}^{\mu}{\cal D}_{j}^{\nu}{\cal D}_{1} \cdot {\cal D}_{2}d 
+ m^{2}~{\cal D}_{i}^{\mu}{\cal D}_{j}^{\nu}d - f^{\mu\nu}_{ij1} + {2 m^{2} \over 3} C^{\mu\nu}_{1}~f^{\prime}_{1}
\label{3.3}
\ee
and another two sets of relations which can be obtained by circular permutations. Here

\bea
f^{\mu\nu}_{221} = f^{\mu\nu}_{331} = A^{\mu\nu}_{1}f_{1},\quad
f^{\mu\nu}_{231} = - B^{\mu\nu}_{1}f_{1},
\nonumber \\
f^{\mu\nu}_{131} = 
(\partial^{\nu}_{1}\partial^{\mu}_{2} + A^{\mu\nu}_{1})f_{1},\quad
f^{\mu\nu}_{121} = 
- (\partial^{\nu}_{1}\partial^{\mu}_{2} + B^{\mu\nu}_{1})f_{1},
\nonumber \\
f^{\mu\nu}_{111} = 
(\partial^{\nu}_{1}\partial^{\mu}_{2} + \partial^{\mu}_{1}\partial^{\nu}_{2}
+ 2 \partial^{\mu}_{2}\partial^{\nu}_{2} + A^{\mu\nu}_{1})f_{1}
\eea
and the rest by circular permutations. Here we have defined
\bea
A^{\mu\nu}_{j} \equiv {2 \over 3} \left(\partial^{\mu}_{j}\partial^{\mu}_{j}
- {1\over 4} \eta^{\mu\nu}~\square_{j}\right)
\nonumber \\
B^{\mu\nu}_{j} \equiv {1 \over 3} \left(\partial^{\mu}_{j}\partial^{\mu}_{j}
+ {1\over 2} \eta^{\mu\nu}~\square_{j}\right)
\nonumber \\
C^{\mu\nu}_{j} \equiv (\partial^{\mu}_{j}\partial^{\mu}_{j}
- \eta^{\mu\nu}~\square_{j}).
\eea

After the central causal splitting we obtain
\be
{\partial \over \partial x^{\rho}} ({\cal D}_{i}^{\mu}{\cal D}_{j}^{\nu}{\cal D}_{1}^{\rho}d)^{F}
= ({\cal D}_{i}^{\mu}{\cal D}_{j}^{\nu}{\cal D}_{1} \cdot {\cal D}_{3}d)^{F} 
- ({\cal D}_{i}^{\mu}{\cal D}_{j}^{\nu}{\cal D}_{1} \cdot {\cal D}_{2}d)^{F}
\label{3.1F}
\ee
\be
{\partial \over \partial y^{\rho}} ({\cal D}_{i}^{\mu}{\cal D}_{j}^{\nu}{\cal D}_{1}^{\rho}d)^{F} = 
-  ({\cal D}_{i}^{\mu}{\cal D}_{j}^{\nu}{\cal D}_{1} \cdot {\cal D}_{3}d)^{F} 
- m^{2}~({\cal D}_{i}^{\mu}{\cal D}_{j}^{\nu}d)^{F} + f^{\mu\nu, F}_{ij1} - {2 m^{2} \over 3} C^{\mu\nu}_{1}~f^{\prime,F}_{1}
+ A^{\mu\nu}_{ij1},
\label{3.2F}
\ee
\be
{\partial \over \partial z^{\rho}} ({\cal D}_{i}^{\mu}{\cal D}_{j}^{\nu}{\cal D}_{1}^{\rho}d)^{F} 
= ({\cal D}_{i}^{\mu}{\cal D}_{j}^{\nu}{\cal D}_{1} \cdot {\cal D}_{2}d)^{F} 
+ m^{2}~({\cal D}_{i}^{\mu}{\cal D}_{j}^{\nu}d)^{F} - f^{\mu\nu,F}_{ij1} + {2 m^{2} \over 3} C^{\mu\nu}_{1}~f^{\prime,F}_{1}
- A^{\mu\nu}_{ij1}
\label{3.3F}
\ee
and another two sets of relations which can be obtained by circular permutations. Here
\be
A^{\mu\nu}_{ijk} \equiv C~\left( a_{jk}^{\mu\nu} + {2 \over 3} C_{l}^{\mu\nu}\right)~\delta
\ee

Here we have defined the differential operators
\bea
a_{11}^{\mu\nu} \equiv \partial_{2}^{\mu}\partial_{2}^{\nu}
+ \partial_{3}^{\mu}\partial_{3}^{\nu}
- {1\over 2} (\partial_{2}^{\mu}\partial_{3}^{\nu}
+ \partial_{3}^{\mu}\partial_{2}^{\nu})
- {1\over 2}~\eta^{\mu\nu}~(\square_{2} + \square_{3} 
+ \partial_{2}\cdot\partial_{3})
\nonumber \\
a_{12}^{\mu\nu} \equiv - \partial_{1}^{\mu}\partial_{1}^{\nu}
- \partial_{2}^{\mu}\partial_{2}^{\nu}
- {1\over 2} (\partial_{1}^{\mu}\partial_{2}^{\nu}
+ \partial_{2}^{\mu}\partial_{1}^{\nu})
- {1\over 2}~\eta^{\mu\nu}~(\square_{1} + \square_{2} 
+ \partial_{1}\cdot\partial_{2})
\eea
and 
$
a_{22}^{\mu\nu}, a_{33}^{\mu\nu}, a_{23}^{\mu\nu}, a_{31}^{\mu\nu}
$
by circular permutations and
$
C = {1 \over 6}~A.
$
\label{split3}
\end{thm}
{\bf Proof:}
We consider the relation (\ref{3.2}). The anomaly is, in momentum space:
\be
\tilde{\cal A}_{ij1}^{\mu\nu} \equiv 
- i~q_{\rho} \tilde{a}^{\mu\nu\rho}_{ij1} + \tilde{a}^{\mu\nu}_{ij13}
+ m^{2}~\tilde{a}^{\mu\nu}_{ij} - (\tilde{f}_{ij1}^{\mu\nu})^{\rm adv}
- {2 m^{2} \over 3}~\tilde{C}_{1}^{\mu\nu}(\tilde{f}^{\prime})^{\rm adv}_{1}
\ee
where 
$
\tilde{C}^{\mu\nu}_{1}
$
is obtained from 
$
C^{\mu\nu}_{1}
$
making 
$
\partial_{1} \rightarrow p, \partial_{2} \rightarrow q.
$
By the same mechanism as before we have:
\be
\tilde{\cal A}_{ij1}^{\mu\nu}(p,q) = - {i m^{2} \over 2\pi} \int {dt\over t^{3}} (1 + t)
\tilde{d}^{\mu\nu}_{ij}(tp, tq)
+ {2i m^{2}\over 6\pi} \tilde{C}^{\mu\nu}_{1}\int {dt\over t} \tilde{f}^{\prime}_{1}(tp, tq).
\ee
If we use (\ref{d21}) we obtain:
\bea
\tilde{\cal A}_{ij1}^{\mu\nu}(p,q) =
{i m^{2}\over 2 \pi}~p^{\mu}p^{\nu}~\int {dt\over t}\tilde{A}_{ij}(tp,tq) 
+ {i m^{2}\over 2 \pi}~q^{\mu}q^{\nu}~\int {dt\over t}\tilde{B}_{ij}(tp,tq) 
\nonumber \\
+ {i m^{2}\over 2 \pi}~p^{\mu}q^{\nu}~
\int {dt\over t}\tilde{C}^{(1)}_{ij}(tp,tq)
+ {i m^{2}\over 2 \pi}~q^{\mu}p^{\nu}~
\int {dt\over t}\tilde{C}^{(2)}_{ij}(tp,tq) 
\nonumber \\
-\eta^{\mu\nu}~{i m^{2}\over 2 \pi}~
\int {dt\over t^{3}}\tilde{D}_{ij}(tp,tq)
+ {i m^{2}\over 3 \pi}~(p^{\mu} p^{\nu} - \eta^{\mu\nu} p^{2})
\int {dt\over t} \tilde{f}_{1}^{\prime}(tp,tq).
\eea
If we substitute the formulas for the functions
$
\tilde{A}_{jk}(p,q)
$,
etc. obtained previously then we need a few more integrals; the first is:
\be
a^{\prime} \equiv \int {dt\over t^{3}} \tilde{d}_{m,m,m}(tp, tq).
\label{a-prime}
\ee
Proceeding as in \cite{Sc1} we obtain
\be
a^{\prime} = {b \over m^{2}}~(p^{2} + q^{2} + p \cdot q).
\label{a-prime1}
\ee
Finally we need
\be
\int {dt\over t} \tilde{f}^{\prime}_{j}(tp, tq) = b.
\label{f-prime}
\ee
Using all these formulas we obtain the result from the statement.
$\qed$

\newpage

Now we have relations similar to those from the previous theorems for the one-particle reducible distributions of the type
(\ref{df}). 
\begin{thm}
The following relations are true
\be
{\partial \over \partial y^{\rho}}{\cal D}^{\rho}_{2}d^{(3)} = - m^{2}d^{(3)} - f_{2},
\label{shell-1PR-1}
\ee
\be
{\partial \over \partial y^{\rho}}{\cal D}^{\mu}_{1}{\cal D}^{\rho}_{2}d^{(3)} = - m^{2}{\cal D}^{\mu}_{1}d^{(3)} + \partial_{2}^{\mu}f_{2},
\label{shell-1PR-2}
\ee
\be
{\partial \over \partial y^{\rho}}{\cal D}^{\mu}_{2}{\cal D}^{\rho}_{2}d^{(3)} = - m^{2}{\cal D}^{\mu}_{2}d^{(3)} - \partial_{1}^{\mu}f_{2},
\label{shell-1PR-3}
\ee
\be
{\partial \over \partial y^{\rho}}{\cal D}^{\mu}_{1}{\cal D}^{\nu}_{1}{\cal D}^{\rho}_{2}d^{(3)} =
- m^{2}{\cal D}^{\mu}_{1}{\cal D}^{\nu}_{1}d^{(3)} + \partial_{2}^{\mu}\partial_{2}^{\nu}f_{2},
\label{shell-1PR-4}
\ee
\be
{\partial \over \partial y^{\rho}}{\cal D}^{\mu}_{2}{\cal D}^{\nu}_{2}{\cal D}^{\rho}_{2}d^{(3)} =
- m^{2}{\cal D}^{\mu}_{2}{\cal D}^{\nu}_{2}d^{(3)} + \partial_{1}^{\mu}\partial_{1}^{\nu}f_{3},
\label{shell-1PR-5}
\ee
\be
{\partial \over \partial y^{\rho}}{\cal D}^{\mu}_{1}{\cal D}^{\nu}_{2}{\cal D}^{\rho}_{2}d^{(3)} =
- m^{2}{\cal D}^{\mu}_{1}{\cal D}^{\nu}_{2}d^{(3)}  - \partial_{2}^{\mu}\partial_{1}^{\nu}f_{3}
\label{shell-1PR-6}
\ee
and similar relations for the other five distributions of this type.
These relations can be causality split without anomalies.
\label{shell-1PR}
\end{thm}

{\bf Proof:} We can proceed as in the proceeding theorems but there is a simple
way, already noticed before: see (\ref{split4}) and (\ref{split5}).
$\qed$

\begin{rem}
Based on previous experience, for instance (\ref{KG}) versus (\ref{split-KG}) or theorem \ref{split1}, etc. we might be inclined to
think that the origin of the anomalies if the presence of mass factors multiplying distributions of lower order of singularity
as the rest of the equations. However, the preceding theorem is a counter-example to this idea. This point shows how difficult
is to decide a priori which differential equations involving causal distributions will produce anomalies. 
\end{rem}

\newpage
\section{Anomalies in the Third Order of the Perturbation Theory}

\subsection{Tree Anomalies}

We have mentioned in the first subsection of the previous section that we have third order anomalies of tree type. These anomalies
can be obtained as the tree anomalies from the second order of the perturbation theory. 
\begin{thm}
Let us consider the causal commutators
$
D^{N}_{(0)}(T^{I}(x),T^{J}(y),T^{K}(z))
$
and perform the causal splitting, i.e. we obtain the chronological products
$
T^{N}_{(0)}(T^{I}(x),T^{J}(y),T^{K}(z))
$
by making
$
D(x) \rightarrow D^{F}(x)
$
as in the second order of the perturbation theory. Then we have the anomalies 
\be
{\cal A}^{IJK}(x,y,z) \equiv sT^{N}_{(0)}(T^{I}(x),T^{J}(y),T^{K}(z)).
\ee
Only in the case
$
I = J = K = \emptyset
$
the anomaly is non-trivial, namely
\be
{\cal A}^{\emptyset\emptyset\emptyset}(x,y,z) = \delta(x - z)~\delta(y - z)~W(z) + \cdots
\ee
where
\be
W = - {1 \over 2}~g_{abcp}~f^{\prime}_{dpe}~\Phi_{a}\Phi_{b}\Phi_{a}\Phi_{c}\Phi_{d}u_{e}
\ee
and
$
\cdots
$
are anomalies of lower canonical dimension. So, we do not have anomalies of canonical dimension $5$ iff
\be
{\cal S}_{abcd}~(g_{abcp}~f^{\prime}_{dpe}) = 0.
\label{higgs}
\ee
\end{thm}
{\bf Proof:}
Let us consider the case
$
I = [\mu], J = [\nu], K = \emptyset
$
when we have
\bea
sD^{N}_{(0)}(T^{\mu}(x),T^{\nu}(y),T(z)) = d_{Q}D^{N}_{(0)}(T^{\mu}(x),T^{\nu}(y),T(z))
\nonumber\\
- i~[ \partial_{\rho}^{1}D^{N}_{(0)}(T^{\mu\rho}(x),T^{\nu}(y),T(z)) - ( x \leftrightarrow y, \mu \leftrightarrow \nu ) ]
\nonumber\\
- i~\partial_{\mu}^{3}D^{N}_{(0)}(T^{\mu}(x),T^{\nu}(y),T^{\rho}(z))
\label{sD0}
\eea
and the anomalies are produced by the terms with derivatives from the right hand side. From the general expression
(\ref{Dcausal}) we have
\bea
D^{N}_{(0)}(T^{\mu\rho}(x),T^{\nu}(y),T(z)) = \delta( x - y) [ N^{[\mu\rho][\nu]}(y), T(z)]
\nonumber\\
- \delta( x - z) [ N^{[\mu\rho]\emptyset}(z), T^{\nu}(y)]
- \delta( y - z) [ N^{[\nu]\emptyset}(z), T^{\mu\rho}(x)]
\eea
and we need the contributions with the factor
$
\partial^{\rho}_{1}D
$
from this expression. Only the last term gives such a contribution and in the end we find out:
\be
D^{N}_{(0)}(T^{\mu\rho}(x),T^{\nu}(y),T(z)) = \delta(y - z)~\partial^{\rho}D(x - z)~W_{1}^{\mu\nu}(x,z) + \cdots
\ee
where the Wick polynomial
$
W_{1}^{\mu\nu}(x,z)
$
can be written explicitly and $\cdots$ are the terms without the derivative 
$
\partial^{\rho}_{1}.
$
Similarly
\bea
D^{N}_{(0)}(T^{\mu}(x),T^{\nu}(y),T^{\rho}(z)) = \delta( x - y) [ N^{[\mu][\nu]}(y), T^{\rho}(z)]
\nonumber\\
+ \delta( x - z) [ N^{[\mu][\rho]}(z), T^{\nu}(y)]
- \delta( y - z) [ N^{[\nu][\rho]}(z), T^{\mu}(x)]
\eea
and we need the terms with the factor
$
\partial^{\rho}_{3}D.
$
Only the first term can produce such a combination. In the end we get
\be
D^{N}_{(0)}(T^{\mu}(x),T^{\nu}(y),T^{\rho}(z)) = \delta(x - y)~\partial^{\rho}D(x - z)~W_{2}^{\mu\nu}(y,z) + \cdots
\ee
where the Wick polynomial
$
W_{2}^{\mu\nu}(x,z)
$
can be written explicitly. The relation
\bea
sD^{N}_{(0)}(T^{\mu}(x),T^{\nu}(y),T(z)) = 0
\eea
is true because we use the Klein-Gordon equation, like in the second order of perturbation theory. If we make the causal decomposition, then
we get the anomaly
\be
{\cal A}^{[\mu][\nu]\emptyset}(x,y,z) \equiv sT^{N}_{(0)}(T^{\mu}(x),T^{\nu}(y),T(z))
= - i~\delta (x - z)~\delta(y - z)~W^{\mu\nu}(z)
\ee
where 
\be
W^{\mu\nu}(z) \equiv W_{1}^{\mu\nu}(z,z) - W_{1}^{\nu\mu}(z,z) - W_{2}^{\mu\nu}(z,z)
\ee
corresponding to the three terms from the right hand side of (\ref{sD0}). An explicit computation gives
$
W^{\mu\nu} = 0
$
so the anomaly is null. The other cases are considered similarly.
$\qed$

As in the second order of the perturbation theory, we can derive the preceding result using the off-shell method \cite{caciulata4}.
In the particular case of the standard model, from the relation (\ref{higgs}) one can obtain the usual form of the Higgs coupling \cite{Sc2}.
\newpage
\subsection{Loop Anomalies}

We need some definitions. In the Yang-Mills sector we need 
\be
f^{(0)}_{[abc]} = f_{eap} f_{ebq} f_{cpq}
\ee
and 
\be
A_{abc} \equiv \sum_{\epsilon}~\epsilon~Tr(\{t^{\epsilon}_{a}, t^{\epsilon}_{b}\} t^{\epsilon}_{c}).
\ee

In the scalar sector we will need:
\bea
f^{(1)}_{abc} = f^{\prime}_{pae} f^{\prime}_{qbe} f^{\prime}_{pqc},\quad
\nonumber\\
f^{(2)}_{abc} = f^{\prime}_{eap} f^{\prime}_{ebq} f_{cpq} = {1\over 2}
f^{\prime}_{abd}~g_{cd}
\quad
f^{(3)}_{[abc]} = f^{\prime}_{epa} f^{\prime}_{eqb} f^{\prime}_{pqc}
\eea
and in the Dirac sector
\be
f^{(4)}_{[abc]} = i~Tr([ t_{a}^{\epsilon},t_{b}^{\epsilon}]t_{c}^{\epsilon}) =
f_{abd}~g^{(2)}_{cd}
\ee
and
\be
t^{(2)}_{a\epsilon} = \sum_{b}
t^{\epsilon}_{b}t^{\epsilon}_{a}t^{\epsilon}_{b}.
\ee
It is also useful to denote
\be
F_{abc} \equiv - {4 C\over 3}~(7 f^{(0)}_{abc} + 2 f^{(3)}_{abc} + 4 f^{(4)}_{abc} ).
\ee
We have the following result. 
\begin{thm}
Let us perform the central causal splitting for all distributions appearing in the third order causal products. 
Then we obtain the following anomalies:
\be
sT(T^{I}(x),T^{J}(y),T^{K}(z)) = {\cal A}^{IJK}(x,y,z)
\ee
where:

(a) In the Yang-Mills sector we have 

- the even part:
\bea
{\cal A}^{[\mu][\nu]\emptyset}_{\rm even}(x,y,z) =
[\partial_{1}^{\mu}\partial_{1}^{\nu} - \partial_{2}^{\mu}\partial_{2}^{\nu}
- \eta^{\mu\nu}(\square_{1} - \square_{2})]~\delta(x,y,z)~F_{abc}~u_{a}(x) u_{b}(y) u_{c}(z) 
\eea
\bea
{\cal A}^{\emptyset\emptyset[\mu\nu]}_{\rm even} = 0
\eea
\bea
{\cal A}^{\emptyset\emptyset[\mu]}_{\rm even}(x,y,z) = \Bigl\{ [\partial_{2}^{\mu}\partial_{2}^{\nu} 
+ \partial_{1}^{\mu}\partial_{2}^{\nu} + \partial_{1}^{\nu}\partial_{2}^{\mu} 
- \eta^{\mu\nu}(\square_{2} + 2 \partial_{1}\cdot \partial_{2})]~\delta(x,y,z)
\nonumber \\
F_{abc}~v_{a\nu}(x) u_{b}(y) u_{c}(z)
\nonumber\\
+ B(\partial_{1} + 2 \partial_{2})_{\nu}~\delta(x,y,z)~f^{(0)}_{abc}~F_{a}^{\mu\nu}(x) u_{b}(y) u_{c}(z) \Bigl\}
+ ( x \leftrightarrow y) ]
\eea
and
\be
{\cal A}^{\emptyset\emptyset\emptyset]}_{\rm even}(x,y,z) = {\cal A}^{\emptyset\emptyset\emptyset}_{(3,YM)}(x,y,z)
+ ( x \leftrightarrow z) + ( y \leftrightarrow z) 
\ee
where
\bea
{\cal A}^{\emptyset\emptyset\emptyset}_{(3,YM)}(x,y,z) \equiv [\partial_{1}^{\mu}\partial_{1}^{\nu} - \partial_{2}^{\mu}\partial_{2}^{\nu}
- \eta^{\mu\nu}(\square_{1} - \square_{2})]~\delta(x,y,z)
%\nonumber \\
F_{abc}~v_{a\nu}(x) v_{b\nu}(y) u_{c}(z) 
\nonumber\\
+ B [ (\partial_{1}^{\rho} + 2 \partial_{2}^{\rho})\delta(x,y,z)~f^{(0)}_{abc}~
F_{a\rho\sigma}(x) v_{b}^{\sigma}(y) u_{c}(z) + ( x \leftrightarrow y)]
\eea

- In the odd part:

\bea
{\cal A}^{[\mu][\nu]\emptyset}_{\rm odd}(x,y,z)
= - 8 i C~\varepsilon^{\mu\nu\rho\sigma} \partial_{1\rho}\partial_{2\sigma}
~\delta(x,y,z)~A_{abc}~u_{a}(x) u_{b}(y) u_{c}(z)
\nonumber \\
{\cal A}^{\emptyset\emptyset[\mu]}_{\rm odd}(x,y,z)
= - 8 i C~\varepsilon^{\mu\nu\rho\sigma} \partial_{1\rho}\partial_{2\sigma}
~\delta(x,y,z)~A_{abc}~u_{a}(x) v_{b\nu}(y) u_{c}(z) + ( x \leftrightarrow y)
\nonumber \\
{\cal A}^{\emptyset\emptyset\emptyset}_{\rm odd}(x,y,z)
= - 8 i C~\varepsilon^{\mu\nu\rho\sigma} \partial_{1\rho}\partial_{2\sigma}
~\delta(x,y,z)~A_{abc}~v_{a\mu}(x) v_{b\nu}(y) u_{c}(z) 
\nonumber \\
+  ( x \leftrightarrow z) +  ( y \leftrightarrow z)
\eea

(b) In the scalar sector we have only an even part. The non-zero contributions appears only in

\bea
{\cal A}^{\emptyset\emptyset\emptyset}_{(3, {\rm scalar})}(x,y,z) =
- f^{(1)}_{abc} \{ [ 2 B (\partial_{1}^{\mu} + 2 \partial_{2}^{\mu})\delta(x,y,z)
[ \partial_{\mu}\Phi_{a}(x) \Phi_{b}(y) u_{c}(z) + ( x \leftrightarrow y)]
\nonumber\\
- 2 C~(\square_{1} - \square_{2})\delta(x,y,z)~\Phi_{a}(x) \Phi_{b}(y) u_{c}(z) \}
\nonumber \\
- f^{(2)}_{abc} [ - B (2 \partial_{1}^{\mu} + \partial_{2}^{\mu})\delta(x,y,z)~
u_{c}(x) \partial_{\mu}\Phi_{a}(y) \Phi_{b}(z)
\nonumber\\
+ 2 (\square_{1} - \square_{2})\delta(x,y,z)~u_{c}(x) \Phi_{a}(y) \Phi_{b}(z) + ( x \leftrightarrow y)]
\eea
(c) The Dirac sector has even and odd sectors grouped as follows:
\be
{\cal A}^{\emptyset\emptyset\emptyset}_{(3, {\rm Dirac})}(x,y,z)
= 4 B [ (2 \partial_{1}^{\mu} + \partial_{2}^{\mu})\delta(x,y,z)~
u_{a}(z) \bar{\psi}(x) t^{(2)}_{a\epsilon} \otimes \gamma_{\mu}\gamma_{\epsilon} \psi(y) + ( x \leftrightarrow y) ]
\ee
\end{thm}
{\bf Proof:} By definition
\bea
sT(T^{\mu}(x),T^{\nu}(y),T(z)) = d_{Q}T(T^{\mu}(x),T^{\nu}(y),T(z)) 
\nonumber\\
- i~[ \partial_{\rho}^{1}T(T^{\mu\rho}(x),T^{\nu}(y),T(z)) + ( x \leftrightarrow y) ]
\nonumber\\
- i\partial_{\rho}^{3}T(T^{\mu}(x),T^{\nu}(y),T^{\rho}(z))
\label{sTmunu}
\eea
Let us investigate the anomalies produced by the derivative terms in the right hand side. For simplicity we consider 
only the Yang-Mills sector. We must use formula (\ref{D3}).
We remind again the origin of the anomalies. To prove 
$
sD(T^{\mu}(x),T^{\nu}(y),T(z)) = 0
$
we must use the first three relations of theorem \ref{split3}. However, after we perform the central causal splitting
$
D(A(x),B(y),C(z)) \rightarrow T(A(x),B(y),C(z))
$
we obtain 
anomalies according to (\ref{3.1F}) - (\ref{3.3F}). The anomaly produced by the last term of the relation (\ref{sTmunu}) is
\bea
{\cal A}_{1,YM}^{[\mu][\nu]\emptyset}(x,y,z) = [
A_{112}^{\mu\nu} - A_{221}^{\mu\nu} - A_{311}^{\nu\mu} + A_{232}^{\nu\mu} - A_{112}^{\nu\mu} + A_{122}^{\nu\mu}
\nonumber\\
- A_{311}^{\mu\nu} - A_{331}^{\mu\nu} + A_{232}^{\mu\nu} + A_{332}^{\mu\nu} + 2 (A_{312}^{\nu\mu} - A_{231}^{\nu\mu})
\nonumber\\
+ \eta^{\mu\nu} \eta_{\rho\sigma} ( - A_{121}^{\rho\sigma} + A_{122}^{\rho\sigma}) ]~f^{(0)}_{abc}~u_{a}(x)~u_{b}(y)~u_{c}(z)
\eea
In the same way we obtain anomalies from the terms
$
\partial_{\rho}^{1}T(T^{\mu\rho}(x),T^{\nu}(y),T(z)) + ( x \leftrightarrow y)
$
so in the end, we obtain in the Yang-Mills sector the anomaly:
\bea
{\cal A}_{1,YM}^{[\mu][\nu]\emptyset}(x,y,z) = C
\Bigl\{ \Bigl[ a_{11}^{\mu\nu} - a_{12}^{\mu\nu} + a_{12}^{\nu\mu} + a_{23}^{\nu\mu} - {16 \over 3} C_{1}^{\mu\nu} 
%\nonumber\\
- \eta^{\mu\nu} \eta_{\rho\sigma} \Bigl( a_{23}^{\rho\sigma} + {2 \over 3} C_{1}^{\rho\sigma}\Bigl) \Bigl]~~\delta(X)~\delta(Y)
\nonumber\\
- ( x \leftrightarrow y, \mu \leftrightarrow \nu) \Bigl\}~ f^{(0)}_{abc}~u_{a}(x)~u_{b}(y)~u_{c}(z)~~~
\eea
The end result is
\bea
{\cal A}_{YM}^{[\mu][\nu]\emptyset}(x,y,z) =
- {28 C\over 3}~[\partial_{1}^{\mu}\partial_{1}^{\nu} - \partial_{2}^{\mu}\partial_{2}^{\nu}
- \eta^{\mu\nu}(\square_{1} - \square_{2})]~\delta(X)~\delta(Y)
\nonumber \\
f^{(0)}_{abc}~u_{a}(x) u_{b}(y) u_{c}(z)
\eea
The scalar and Dirac contributions can be computed in the same way and we get the first formula from the statement.
The other two formulas are obtained similarly.
$\qed$

The preceding expressions are not unique because of the presence of the delta distribution
$
\delta = \delta(x - z)~\delta(y - z).
$
We can re-express the anomalies in an unique form of the type
$
p(\partial_{1},\partial_{2})\delta~W(z).
$
%\newpage

\begin{thm}
The anomalies
$
{\cal A}^{IJK}(x,y,z)
$
can be uniquely written as follows:

(a) In the Yang-Mills sector we have 

- the even part:
\bea
{\cal A}^{[\mu][\nu]\emptyset}_{\rm even}(x,y,z) = F_{abc}~
[(\partial_{1}^{\mu}\partial_{1}^{\nu} - \eta^{\mu\nu} \square_{1})\delta(x,y,z)~(u_{a}u_{b}u_{c})(z) 
\nonumber\\
- \partial^{\mu}_{1}\delta(x,y,z)~(\partial^{\nu}u_{a}u_{b}u_{c})(z) - \partial^{\nu}_{1}\delta(x,y,z)~(\partial^{\mu}u_{a}u_{b}u_{c})(z)
\nonumber\\
+ 2~\eta^{\mu\nu}~\partial_{\rho}^{1}\delta(x,y,z)~(\partial^{\rho}u_{a}u_{b}u_{c})(z) ]
\nonumber\\
- (x \leftrightarrow y, \mu \leftrightarrow \nu)
\eea
\bea
{\cal A}^{\emptyset\emptyset[\mu\nu]}_{\rm even} = 0
\eea
\bea
{\cal A}^{\emptyset\emptyset[\mu]}_{\rm even}(x,y,z) = F_{abc}~\{
[\partial_{1}^{\mu}\partial_{1}^{\nu} + \partial_{2}^{\mu}\partial_{2}^{\nu} + 
2~(\partial_{1}^{\mu}\partial_{2}^{\nu} + \partial_{1}^{\nu}\partial_{2}^{\mu}) 
\nonumber\\
- \eta^{\mu\nu}(\square_{1} + \square_{2} + 4 \partial_{1}\cdot \partial_{2})]\delta(x,y,z)~(v_{a\nu}u_{b}u_{c})(z)
\nonumber\\
- (\partial^{\mu}_{1} + \partial^{\mu}_{2})\delta(x,y,z)~(2 v_{a}^{\nu}\partial_{\nu}u_{b}u_{c} + \partial_{\nu}v_{a}^{\nu}u_{b}u_{c})(z)  
\nonumber\\
+ (\partial^{\nu}_{1} + \partial^{\nu}_{2})\delta(x,y,z)~(2 v_{a}^{\mu}\partial^{\nu}u_{b}u_{c} 
- \partial^{\mu}v_{a}^{\nu}u_{b}u_{c} + 2\partial^{\nu}v_{a}^{\mu}u_{b}u_{c} )(z)  
\nonumber\\
+ 2~\delta(x,y,z)~( \partial^{\mu}v_{a}^{\nu}\partial_{\nu}u_{b}u_{c} - 2\partial^{\nu}v_{a}^{\mu}\partial_{\nu}u_{b}u_{c} 
+ \partial_{\nu}v_{a}^{\nu}\partial^{\mu}u_{b}u_{c} + v_{a}^{\nu}\partial^{\mu}\partial_{\nu}u_{b}u_{c})(z) \}
\nonumber\\
+ B~f^{(0)}_{abc}~[ 3 (\partial^{1}_{\nu} + \partial^{2}_{\nu})\delta(x,y,z)~(F_{a}^{\mu\nu}u_{b}u_{c})(z)
 - 2 \delta(x,y,z)~( \partial^{\mu}\partial_{\nu}v_{a}^{\nu}u_{b}u_{c} + 2F_{a}^{\mu\nu}\partial_{\nu}u_{b}u_{c})(z) ]
\eea
and
\bea
{\cal A}^{\emptyset\emptyset\emptyset}(x,y,z) = 2 \delta(x,y,z) [ F_{abc}( \partial^{\mu}\partial_{\nu}v_{a}^{\nu}v^{\mu}_{b}u_{c}
- \partial_{\nu}v_{a}^{\mu}v^{\nu}_{b}\partial_{\mu}u_{c} + 2 \partial^{\mu}v_{a}^{\nu}v_{b\nu}\partial_{\mu}u_{c}
- \partial_{\nu}v_{a}^{\nu}v^{\mu}_{b}\partial_{\mu}u_{c})
\nonumber\\
+ 3 B f^{(0)}_{abc}~(F_{a}^{\mu\nu}v_{b\mu}\partial_{\nu}u_{c})](z)
\eea

- In the odd part:

\bea
{\cal A}^{[\mu][\nu]\emptyset}_{\rm odd}(x,y,z)
= - 8 i C~A_{abc}~\delta(x,y,z)~\varepsilon^{\mu\nu\rho\sigma}~(\partial_{\rho}u_{a}\partial_{\sigma}u_{b}u_{c})(z)
\nonumber \\
{\cal A}^{\emptyset\emptyset[\mu]}_{\rm odd}(x,y,z)
= 8 i C~A_{abc}~\varepsilon^{\mu\nu\rho\sigma} 
[ (\partial^{1}_{\nu} + \partial^{2}_{\nu})\delta(x,y,z)~(\partial_{\rho}u_{a}v_{b\sigma}u_{c})(z)
\nonumber\\
+ \delta(x,y,z)~(\partial_{\nu}u_{a}F_{b\rho\sigma}u_{c})(z)]
\nonumber \\
{\cal A}^{\emptyset\emptyset\emptyset}_{\rm odd}(x,y,z)
= - 8 i C~A_{abc}~\varepsilon_{\mu\nu\rho\sigma}~\delta(x,y,z)~\left(  - {1 \over 4}F_{a}^{\mu\nu}F_{b}^{\rho\sigma}u_{c}
+ \partial^{\mu}u_{a}v_{b}^{\nu}F_{c}^{\rho\sigma}\right)(z)
\eea

(b) In the scalar sector 
\be
{\cal A}^{\emptyset\emptyset\emptyset}_{\rm scalar}(x,y,z) =
( - 2 C~f^{(1)}_{abc} + 3 B~f^{(2)}_{abc})~\delta(x,y,z)~ (\partial^{\mu}\Phi_{a}\Phi_{b}\partial_{\mu}u_{c})(z)
\ee
(c) In the Dirac sector:
\be
{\cal A}^{\emptyset\emptyset\emptyset}_{\rm Dirac)}(x,y,z) =
- 24 B~\delta(x,y,z)~(\partial^{\mu}u_{a} \bar{\psi} t^{(2)}_{a\epsilon} \otimes \gamma_{\mu}\gamma_{\epsilon} \psi)(z)
\ee
\end{thm}
%\newpage

The nest task is to investigate if the preceding anomalies can be eliminated by a redefinition of the chronological
products. This can be done {\it iff} the anomalies can be written as a coboundary i.e.
\be
{\cal A}^{IJK} = (sB)^{IJK} = d_{Q}B^{IJK} - i (\delta B)^{IJK}
\label{coboundary}
\ee
where the expressions
$
B^{IJK}
$
are quasi-local, Lorentz covariant, of canonical dimension $4$ and with the same (graded) symmetry in 
$
(x,I), (y,J), (z,K)
$
as the chronological products - see (\ref{sqew}). We will write them in the unique form 
$
p(\partial_{1},\partial_{2})\delta~W(z)
$
used in the previous theorem.
\begin{thm}
The generic form of the coboundaries:

(a) In the Yang-Mills sector

- the even part with respect to parity:
\be
B^{\emptyset\emptyset[\mu\nu\rho]} = 0
\ee
\bea
B^{[\mu][\nu][\rho]}(x,y,z) = 
\nonumber\\
k_{abc}~\{ [\eta^{\mu\nu}~(\partial^{\rho}_{1} - \partial^{\rho}_{2}) 
- \eta^{\mu\rho}~(2 \partial^{\nu}_{1} + \partial^{\nu}_{2})
+ \eta^{\nu\rho}~(\partial^{\mu}_{1} + 2 \partial^{\mu}_{2}) ]\delta(x,y,z)~(u_{a}u_{b}u_{c})(z)
\nonumber\\
+ 3~\delta(x,y,z)~(\eta^{\mu\rho}~\partial^{\nu}u_{a}u_{b}u_{c} - \eta^{\nu\rho}~\partial^{\mu}u_{a}u_{b}u_{c})(z) \}
\eea
\bea
B^{[\mu\nu][\rho]\emptyset}(x,y,z) = 
\nonumber\\
\eta^{\mu\rho}~[ p^{1}_{abc}~\partial^{\nu}_{1}\delta(x,y,z)~(u_{a}u_{b}u_{c})(z)
+ p^{2}_{abc}~\partial^{\nu}_{2}\delta(x,y,z)~(u_{a}u_{b}u_{c})(z)
\nonumber\\
+ p^{3}_{abc}~\delta(x,y,z)~(\partial^{\nu}u_{a}u_{b}u_{c})(z) ] - (\mu \leftrightarrow \nu)
\eea
\bea
B^{[\mu][\nu]\emptyset}(x,y,z) = 
%\nonumber\\
q^{1}_{abc}~[\partial^{\mu}_{1}\delta(x,y,z)~(v^{\nu}_{a}u_{b}u_{c})(z)
- \partial^{\nu}_{2}\delta(x,y,z)~(v^{\mu}_{a}u_{b}u_{c})(z)]
\nonumber\\
+ q^{2}_{abc}~[\partial^{\nu}_{1}\delta(x,y,z)~(v^{\mu}_{a}u_{b}u_{c})(z)
- \partial^{\mu}_{2}\delta(x,y,z)~(v^{\nu}_{a}u_{b}u_{c})(z)]
\nonumber\\
+ q^{3}_{abc}~\eta^{\mu\nu}~(\partial^{\rho}_{1} - \partial^{\rho}_{2}) \delta(x,y,z)~(v^{\rho}_{a}u_{b}u_{c})(z)
\nonumber\\
+ q^{4}_{abc}~\delta(x,y,z)~(v^{\mu}_{a}\partial^{\nu}u_{b}u_{c} - v^{\nu}_{a}\partial^{\mu}u_{b}u_{c})(z) 
\nonumber\\
+ q^{5}_{abc}~\delta(x,y,z)~(F^{\mu\nu}_{a}\partial^{\nu}u_{b}u_{c})(z)
\eea
\bea
B^{\emptyset\emptyset[\mu\nu]}(x,y,z) = 
%\nonumber\\
r^{1}_{abc}~[(\partial^{\mu}_{1} + \partial^{\mu}_{2})\delta(x,y,z)~(v^{\nu}_{a}u_{b}u_{c})(z)
- (\partial^{\nu}_{1} + \partial^{\nu}_{2})\delta(x,y,z)~(v^{\mu}_{a}u_{b}u_{c})(z)]
\nonumber\\
+ r^{2}_{abc}~\delta(x,y,z)~(v^{\mu}_{a}\partial^{\nu}u_{b}u_{c} - v^{\nu}_{a}\partial^{\mu}u_{b}u_{c})(z)
\nonumber\\
+ r^{3}_{abc}~\delta(x,y,z)~(F^{\mu\nu}_{a}\partial^{\nu}u_{b}u_{c})(z)
\eea
\bea
B^{\emptyset\emptyset[\mu]}(x,y,z) = 
%\nonumber\\
s^{1}_{abc}~(\partial_{\nu}^{1} + \partial_{\nu}^{2})\delta(x,y,z)~(v^{\mu}_{a}v^{\nu}_{b}u_{c})(z)
%\nonumber\\
+ s^{2}_{abc} (\partial^{\mu}_{1} + \partial^{\mu}_{2})\delta(x,y,z)~(u_{a}u_{b}\tilde{u}_{c})(z)
\nonumber\\
+ s^{3}_{abc}~\delta(x,y,z)~(\partial^{\mu}v^{\nu}_{a}v_{b\nu}u_{c})(z)
%\nonumber\\
+ s^{4}_{abc}~\delta(x,y,z)~(\partial_{\nu}v^{\mu}_{a}v_{b}^{\nu}u_{c})(z)
\nonumber\\
+ s^{5}_{abc}~\delta(x,y,z)~(\partial_{\nu}v^{\nu}_{a}v_{b}^{\mu}u_{c})(z)
%\nonumber\\
+ s^{6}_{abc}~\delta(x,y,z)~(v^{\mu}_{a}v_{b}^{\nu}\partial_{\nu}u_{c})(z)
\nonumber\\
+ s^{7}_{abc} \delta(x,y,z)~(\partial^{\mu}u_{a}u_{b}\tilde{u}_{c})(z)
%\nonumber\\
+ s^{8}_{abc} \delta(x,y,z)~(u_{a}u_{b}\partial^{\mu}\tilde{u}_{c})(z)
\nonumber\\
+ s^{9}_{abc}~(\partial^{\mu}_{1} + \partial^{\mu}_{2})\delta(x,y,z)~(v^{\nu}_{a}v_{b\nu}u_{c})(z)
+ s^{10}_{abc}~\delta(x,y,z)~(v^{\nu}_{a}v_{b\nu}\partial^{\mu}u_{c})(z)
\eea
\bea
B^{\emptyset\emptyset\emptyset}(x,y,z) = 
%\nonumber\\
\delta(x,y,z)(t^{1}_{abc}~\partial^{\mu}v^{\nu}_{a}v_{b\mu}v_{c\nu}
%\nonumber\\
+ t^{2}_{abc}~\partial_{\nu}v^{\nu}_{a}u_{b}\tilde{u}_{c}
\nonumber\\
+ t^{3}_{abc}~v^{\mu}_{a}\partial_{\mu}u_{b}\tilde{u}_{c}
%\nonumber\\
+ t^{4}_{abc}~v^{\mu}_{a}u_{b}\partial_{\mu}\tilde{u}_{c}
%\nonumber\\
+ t^{5}_{abc}~\partial_{\mu}v^{\mu}_{a}v_{b}^{\nu}v_{c\nu})(z)
\eea

- the odd part with respect to parity
\be
B^{\emptyset\emptyset[\mu\nu\rho]} = \varepsilon^{\mu\nu\rho\sigma}~
[ d^{1}_{abc}(\partial_{\sigma}^{1} + \partial_{\sigma}^{2})\delta(x,y,z)~(u_{a}u_{b}u_{c})(z)
+ d^{2}_{abc}~\delta(x,y,z)~(\partial_{\sigma}u_{a}u_{b}u_{c})(z) ]
\ee
\be
B^{[\mu][\nu][\rho]}(x,y,z) = e_{abc}~\varepsilon^{\mu\nu\rho\sigma}~\delta(x,y,z)~(\partial_{\sigma}u_{a}u_{b}u_{c})(z)
\ee
\bea
B^{[\mu\nu][\rho]\emptyset}(x,y,z) = \varepsilon^{\mu\nu\rho\sigma}~
[ ( f^{1}_{abc}~\partial_{\sigma}^{1} + f^{2}_{abc}~\partial_{\sigma}^{2})\delta(x,y,z)~(u_{a}u_{b}u_{c})(z)
\nonumber\\
+ f^{3}_{abc}~\delta(x,y,z)~(\partial_{\sigma}u_{a}u_{b}u_{c})(z) ]
\eea
\bea
B^{[\mu][\nu]\emptyset}(x,y,z) = \varepsilon^{\mu\nu\rho\sigma}~
[  g^{1}_{abc}~(\partial_{\rho}^{1} + \partial_{\rho}^{2})\delta(x,y,z)~(v_{a\sigma}u_{b}u_{c})(z)
\nonumber\\
+ \delta(x,y,z)~( g^{2}_{abc}~F_{a\rho\sigma}u_{b}u_{c}
+ g^{3}_{abc}~v_{a\rho}\partial_{\sigma}u_{b}u_{c} )(z) ]
\eea
\bea
B^{\emptyset\emptyset[\mu\nu]}(x,y,z) = \varepsilon^{\mu\nu\rho\sigma}~
[  h^{1}_{abc}~(\partial_{\rho}^{1} + \partial_{\rho}^{2})\delta(x,y,z)~(v_{a\sigma}u_{b}u_{c})(z)
\nonumber\\
+ \delta(x,y,z)~( h^{2}_{abc}~F_{a\rho\sigma}u_{b}u_{c}
+ h^{3}_{abc}~v_{a\rho}\partial_{\sigma}u_{b}u_{c} )(z) ]
\eea
\bea
B^{\emptyset\emptyset[\mu]}(x,y,z) = \varepsilon^{\mu\nu\rho\sigma}~
[ j^{1}_{abc}~(\partial_{\nu}^{1} + \partial_{\nu}^{2})\delta(x,y,z)~(v_{a\rho}v_{b\sigma}u_{c})(z)
\nonumber\\
+ \delta(x,y,z)~(j^{2}_{abc} v_{a\nu}F_{b\rho\sigma}u_{c} + j^{3}_{abc} v_{a\rho}v_{b\sigma}\partial_{\nu}u_{c})(z)]
\eea
\be
B^{\emptyset\emptyset\emptyset}(x,y,z) = \varepsilon^{\mu\nu\rho\sigma}~
\delta(x,y,z)l_{abc}~F_{a\mu\nu}v_{b\rho}v_{c\sigma}
\ee

(b) In the scalar sector we have only an even part:
\be
B^{IJK} = 0, \quad |I| + |J| + |K| = 0,2,3
\ee
\bea
B^{\emptyset\emptyset[\mu]}(x,y,z) = 
%\nonumber\\
w^{1}_{abc}~(\partial^{\mu}_{1} + \partial^{\mu}_{2})\delta(x,y,z)~(\Phi_{a}\Phi_{b}u_{c})(z)
\nonumber\\
+ w^{2}_{abc}~\delta(x,y,z)~(\partial^{\mu}\Phi_{a}\Phi_{b}u_{c})(z)
+ w^{3}_{abc}~\delta(x,y,z)~(\Phi_{a}\Phi_{b}\partial^{\mu}u_{c})(z)
\eea

(c) In the Dirac sector:
\be
B^{IJK} = 0, \quad |I| + |J| + |K| = 2,3
\ee
\bea
B^{\emptyset\emptyset[\mu]}(x,y,z) = \delta(x,y,z)~V^{\mu}(z)
\nonumber\\
B^{\emptyset\emptyset\emptyset}(x,y,z) \delta(x,y,z)~V(z)
\eea
where
\bea
V^{\mu} = u_{a} \bar{\Psi} V_{a\epsilon} \otimes \gamma^{\mu}\gamma_{\epsilon} \Psi
\nonumber\\
V = u_{a} \bar{\Psi} V^{\prime}_{a\epsilon} \otimes \gamma_{\epsilon} \Psi
\eea

In the preceding expressions we can suppose convenient (anti)symmetry properties of the coefficients.
\end{thm}

%\newpage

Now we impose (\ref{coboundary}). In the even sector the anomaly has only the coefficients
$
F_{abc}, f^{(0)}_{abc}, f^{(1)}_{abc}
$
and
$
f^{(2)}_{abc},
$
which are completely antisymmetric in 
$
a,b,c
$
so if we want to prove that we have a solution of the equation (\ref{coboundary}) in this sector, it is sufficient to suppose
that all coefficients
$
k_{abc}, p^{j}_{abc}, q^{j}_{abc}, r^{j}_{abc}, s^{j}_{abc}, t^{j}_{abc}
$
and
$
w^{j}_{abc},
$
are completely antisymmetric in
$
a,b,c.
$
In particular, it means that we can take
$
s^{j}_{abc} = 0,~j = 9, 10,\quad t^{5}_{abc} = 0,\quad  w^{j}_{abc} = 0,~j = 1,3.
$

For simplicity we denote 
$
k = k_{abc}, p_{j} = p^{j}_{abc},
$
etc. and we have from (\ref{coboundary}) the following system: 

(a) In the Yang-Mills sector:

\bea
p_{1} - k = F
\nonumber\\
q_{3} - p_{3} - 3 k = 2 F
\nonumber\\
q_{1} = - F
\nonumber\\
q_{2} + p_{3} + 3 k = - F
\eea
\be
r_{1} - p_{3} - 3 p_{1} = 0
\ee
\bea
q_{1} + q_{2} + r_{1} = F
\nonumber\\
q_{1} + q_{2} + 2 q_{3} - r_{1} = - F
\nonumber\\
q_{1} + q_{3} - r_{1} = - 2 F
\nonumber\\
q_{2} + r_{1} = 2 F
\nonumber\\
q_{3} - r_{1} = - F
\nonumber\\
s_{2} + q_{3} - r_{1} = - F
\nonumber\\
- s_{1} - q_{4} + 2 q_{1} - r_{2} = 0
\nonumber\\
s_{1} + q_{4} + 2 q_{2} + 2 r_{1} + r_{2} = 2 F
\nonumber\\
q_{1} + q_{5} + r_{3} = - F + 3 B~f^{(0)}
\nonumber\\
q_{2} - q_{5} + r_{1} - r_{3} = 2 F - 3 B~f^{(0)}
\nonumber\\
s_{3} + s_{4} - r_{2} = - 2 F
\nonumber\\
s_{3} - 2 r_{3} = 2 F - 4 B~f^{(0)}
\nonumber\\
s_{4} - r_{2} + 2 r_{3} = - 4 F + 4 B~f^{(0)}
\nonumber\\
s_{5} - s_{7} + r_{2} = 2 F
\nonumber\\
s_{6} - r_{2} = 0
\nonumber\\
s_{8} + r_{3} = 2 B~f^{(0)}
\eea
\bea
t_{4} - s_{4} - s_{5} = 2 F
\nonumber\\
- t_{1} - s_{3} = 4 F - 6 B~f^{(0)}
\nonumber\\
t_{1} - s_{4} + s_{6} = - 2 F + 6 B~f^{(0)}
\nonumber\\
t_{3} - s_{5} - s_{6} = - 2 F
\nonumber\\
t_{3} - 4 s_{2} - s_{7} = 0
\nonumber\\
t_{4} - 8 s_{2} - s_{7} - 2 s_{8} = 0
\eea

(b) In the scalar sector:
\bea
w_{2} = 2 C f^{(1)} - B f^{(2)}
\eea

One can prove easily that the preceding system of equations has a solution. A interesting problem is if we really need to
renormalize the expression
$
T(T(x),T(y),T(z))
$
i.e. if we can take
\be
B^{\emptyset\emptyset\emptyset} = 0
\ee
or not. It can be proved that the preceding equality is equivalent to
\be
F_{abc} = B f^{(0)}_{abc}.
\ee

In the odd sector, because the anomaly involves only the coefficient
$
A_{abc}
$
which is completely symmetric in
$
a,b,c
$
we can consider that all coefficients
$
d^{\alpha}_{abc}, e_{abc}, f^{\alpha}_{abc}, g^{\alpha}_{abc}, h^{\alpha}_{abc},j^{\alpha}_{abc}, l_{abc}
$
are completely symmetric in
$
a,b,c
$.

In particular, it means that we can take
$
d^{j}_{abc}, e_{abc}, f^{\alpha}_{abc}, g^{\alpha}_{abc},\alpha = 1,2,~g^{\alpha}_{abc},\alpha = 1,2, j^{\alpha}_{abc},\alpha = 1,3
$
and
$
l_{abc}
$
to be zero, i.e. only the coefficients
$
g^{3}_{abc}, h^{3}_{abc}
$
and
$
j^{2}_{abc}
$
survive. 

It is sufficient to consider only the case 
$
I = J = K = \emptyset
$
of (\ref{coboundary}). We obtain:
\bea
j_{2} = 2 C A
\nonumber\\
- 2 j_{2} = - 8 C A
\eea
which are leading to the equality 
\be
A_{abc} = 0.
\ee

This is exactly the standard form (see for instance \cite{Huang} formula (11.58)) for the cancellation of the axial anomaly.

Finally, in the Dirac sector we have the solution of (\ref{coboundary})
\be
V_{a\epsilon} = - 24 B t^{(2)}_{a\epsilon}.
\ee

\newpage
\section{Conclusions}

In the functional formalism one considers anomalies of the current
conservation 
\be
\partial_{\mu}j^{\mu}_{\rm Axial} - j^{5}_{\rm Axial}
\ee
or of the BRST invariance of the generating functional of the Green distributions
\be
s_{BRST}\Gamma
\ee
(where $s_{BRST}$ is the non-linear BRST operator from the functional formalism).
In this formalism the anomaly has terms cubic and quartic in the fields 
- see for instance \cite{BBBC} formula (13).
The cubic term 
$
\varepsilon^{\mu\nu\rho\sigma} \partial_{1\rho}\partial_{2\sigma}
~\delta(X)~\delta(Y)~A_{abc}~v_{a\mu}(x) v_{b\nu}(y) u_{c}(z) 
$
obtained above coincides with the first contribution of this formula, up to partial integration. 
To obtain the quartic terms one would have to go to the fourth order of the perturbation theory.

We have investigated the anomalies of the standard model of maximal canonical dimension 
$
\omega = 5
$
in the third order of the perturbation theory for tree and one-loop contributions. Anomalies of lower canonical dimension must be investigated 
separately using Wess-Zumino consistency relations (\ref{WZ}). The analysis goes as follows. The dominant
contribution to the anomaly considered in this paper was obtained, essentially, by replacing everywhere
the Pauli-Jordan distributions
$
D_{m_{j}}(x)
$
of various mases by 
$
D_{M}(x)
$
where $M$ is some fixed positive mass. This substitution implies a corresponding splitting of the chronological products. 
The dominant contribution to the chronological products gives the dominant contribution to the anomaly and we have showed 
how this anomaly can be eliminated. It follows that we still have potential anomalies of canonical dimension with $2$
units lower i.e. of maximal canonical dimension $3$. So, a priori, we still might have anomalies of the type
\be
\delta(x - z)\delta(y - z)~W(z)
\ee
with $W$ a Wick polynomial of canonical dimension $3$. But in \cite{cohomology} we have proved that such anomalies are null due to
the Wess-Zumino consistency relations (\ref{WZ}).  We still have to investigate the anomalies associated to two-loops graphs. The analysis is also
cohomological \cite{cohomology}. The anomalies must be of the form
\be
{\cal A}^{IJK}(x,y,z) = p(\partial)\delta(x,y,z)~w(z)
\ee
where $W$ is linear in the fields. Because of the condition
$
gh({\cal A}^{IJK}) = |I| + |J| + |K| + 1
$
only the case
$
I = J = K = \emptyset
$
can produce anomalies. If the polynomial $p$ is non-trivial one can easily exhibit the anomaly in the form of a coboundary. So we are left with
\be
{\cal A}^{\emptyset\emptyset\emptyset}(x,y,z) = \delta(x,y,z)~w(z)
\ee
with 
$
gh(w) = 1
$
i.e.
$
w = \sum_{a} f_{a}~u_{a}.
$
We can write the contributions corresponding to
$
a \in I_{2}
$
as 
$
d_{Q}b
$
so we are left with the case 
$
a \in I_{1}.
$
For $a$ corresponding to gluons we must use the fact that there is no vector
$
f_{a}
$
invariant with respect to 
$
SU(3)
$
and for $a$
corresponding to the photon we can use charge invariance. So there are no anomalies for two-loops graphs in the third order.

The generalization of the preceding analysis to multi-loop contributions in not obvious and it is a
subject of further investigation.

\end{document}